\documentclass[12pt]{amsart}
\usepackage{subfigure}
\usepackage{subfloat}
\usepackage{lineno}
\usepackage{etex}
\usepackage{tikz}
\usetikzlibrary{arrows,shapes}

\usepackage{setspace}
\usepackage{mathrsfs}
\usepackage{mathtools}
\usepackage{subcaption}
\usepackage{fullpage}
\usepackage{graphics}
\usepackage{latexsym}

\usepackage{times}
\usepackage{natbib}
\usepackage{algcompatible}
\newcommand{\RETURN}{\\Return}
\usepackage{algorithm}

\usepackage{graphicx}
\usepackage{amsfonts}
\usepackage{amsmath,amssymb,amscd,amsthm}
\usepackage{epsf}
\usepackage{multirow}
\usepackage{multicol}
\usepackage{pstricks,pst-node,pst-tree}
\usepackage{tabularx}
\usepackage{array}

\def\ci{\perp\!\!\!\perp}

\def\Tet{\hat\theta_2^{EL}}

\DeclareMathOperator*{\argmax}{\arg\!\max}
\newtheorem{exm}{\bfseries{Example}}
\newtheorem{theorem}{\bfseries{Theorem}}

\renewcommand{\pmb}[1]{#1}

\newcommand{\go}{\pmb{\gamma}^{(1)}}

\newcommand{\gt}{\pmb{\gamma}^{(t)}}
\newcommand{\gtp}{\pmb{\gamma}^{(t+1)}}
\newcommand{\mc}[1]{\mathcal{#1}}

\def\spacingset#1{\renewcommand{\baselinestretch}%
{#1}\small\normalsize} \spacingset{1}

\title[]{A Two-step Metropolis Hastings Method for Bayesian Empirical Likelihood Computation with Application to Quantile Regression and Bayesian Model Selection}
\author[]{Sanjay Chaudhuri}
\thanks{The first author gratefully acknowledges partial supports from grants from MOE Singapore, P\&G Singapore and grant 2413491 from NSF-DMS USA.}
\address{Department of Statistics, University of Nebraska-Lincoln}
\email{schaudhuri2@unl.edu}
\author[]{Teng Yin}
\address{DFS Group Ltd., Singapore}
\author[]{Snehashis Chakraborty}
\address{Department of Electrical and Computer Engineering, Texas A\& M University}
\author[]{Rupsa Roy}
\address{Lantrasoft}

\begin{document}

\begin{abstract}
  Empirical likelihood-based methods have been used under the Bayesian framework (BayesEL) in recent times. For statistical inference, these methods require efficient Markov chain Monte Carlo (MCMC) samplers for drawing observations from the parameter posterior distributions. However, the complex, especially non-convex, nature of the empirical likelihood support makes such MCMC algorithms harder to design.
  Such difficulties have restricted the use of BayesEL methods in many applications. In this article, we propose a two-step Metropolis-Hastings algorithm to sample from the BayesEL posteriors. Our proposal uses the current values of suitable subsets of the parameters and the estimating equations determining the underlying empirical likelihood to propose values of the remaining parameters.
  The proposed method is thus suitable for sampling from BayesEL posteriors in many complex problems, especially those with discontinuous estimating equations, e.g., simultaneous quantile regression. Furthermore, the proposed method easily extends to BayesEL model selection through a reversible jump Markov chain Monte Carlo procedure. Several illustrative, real-life applications of our proposed methods are presented.
\end{abstract}
\

\maketitle


\spacingset{1.2} 

\section{Introduction}
In recent years, empirical likelihood \citep{owen1988empirical,qin1994empirical} based procedures have been frequently used under Bayesian framework.  Such procedures specify a statistical model through unbiased estimating equations, without 
requiring a declaration of the data distribution.  The likelihood is estimated from the empirical distribution function computed under constraints imposed by these estimating equations. 
The estimated likelihood is then used to define a posterior.    
The validity of empirical and similar likelihoods for Bayesian inference has been a topic of extensive discussion \citep{monahan1992proper,lazar2003bayesian,fang2006empirical,corcoran_1998}.  
Alternative likelihoods like  Bayesian exponential tilted empirical likelihood (BETEL) \citep{schennach2005bayesian} have been proposed and justified using basic probabilistic arguments.
In recent times, many authors \citep{grendarJudge2009,chib_shin_simoni_2017,zhong_ghosh_2017,zhaoGhoshRaoWu2020, sueishi_2024} have considered asymptotic properties of the posteriors and parameter estimates obtained from such likelihoods.  
Due to its convenience in statistical modelling, the Bayesian empirical likelihood (BayesEL) procedures have seen many applications, such as in
analysing complex survey data \citep{rao2010bayesian},  small area estimation \citep{chaudhuri2011empirical,porter_holan_wickle_2015,chenLiu2019}, quantile regression \citep{yang2012bayesian}, among others. 

The likelihood as well as the posterior in BayesEL procedure is computed numerically for each value of the parameter.  Inferences are drawn using the samples generated from the posterior.  However, efficient sampling of BayesEL posteriors require bespoke procedures. 
Metropolis Hastings \citep{hastings1970monte} or similar Markov chain Monte Carlo (MCMC) methods need to be used.  These extensively studied methods, however, are not easily implemented in BayesEL posterior sampling.  The support of the empirical likelihood is data dependent, does not usually cover the whole parameter space and is usually non-convex \citep{chaudhuri_mondal_yin_2017}.  
Proposals which can cover such supports, are not easily constructed. Simple random walks would mix very slowly, since they would often get stuck near the boundary of the support.  
 Methods like the metropolis algorithm in \citet{haario2001adaptive} or parallel tempering \citep{geyer1992markov, liu2008monte} do not adapt to the support non-convexity satisfactorily. 
These difficulties in drawing samples from a BayesEL posterior have been a major impediment to its wider use in statistical modelling. 

Some authors (e.g \citet{mengersenPudloRobert2013}, \citet{porter_holan_wickle_2015}) have designed specific MCMC algorithms to sample from a BayesEL Posterior.  For smooth estimating equations, a more general method was 
proposed by \citet{chaudhuri_mondal_yin_2017}.  They show that the gradient of the log-empirical likelihood diverges at the boundaries of its support, and design a gradient-based Hamiltonian Monte Carlo method to sample from a BayesEL posterior. The diverging gradient ensures that the chain always reflects towards the centre of the support from its boundaries.  Recently, \citet{yuBondel2024} has deduced a variatonal approximation for BayesEL posteriors, which can be used for fast sampling from and approximate posterior density.

If one wishes to explore the possibility of Bayesian model selection using BayesEL procedure, the problem becomes much more acute.   In order to sample from the posterior arising 
in Bayesian model selection problems, reversible jump Markov chain Monte Carlo (RJMCMC) \citep{green1995reversible} sampler which can jump between models is generally used (see  \cite{fan2010reversible,dellaportas2002bayesian,robert2002bayesian}).
The main challenge in RJMCMC is efficient construction of the cross-model proposals. The usual notions that can guide the sampler in the fixed dimensional state space now appear useless. Inefficient proposal makes the reversible jump sampler 
explore the parameter space slowly or even fail entirely. Consequently, the Markov chain takes a long time to converge.  The construction of an efficient proposal 
is a topic of extensively discussion even in fully parametric setup \citep{brooks2003efficient,sisson2005transdimensional,green2009reversible}. Many variants of RJMCMC have also been proposed \citep{al2004improving,jasra2007population}.

Sampling from a BayesEL posterior becomes more difficult when the dimension of the parameter space is allowed to vary between iterations.  When the RJMCMC is used in the BayesEL procedures, constructing an efficient proposal becomes even harder. 
 This is because even for the parameters common between the current and proposed models, the posterior supports may be entirely different. Therefore, for RJMCMC in the BayesEL procedure, 
it is a great challenge to design a cross-model proposal  which can ensure the proposed candidates to be in their new marginal supports.

In this work, we propose a two-step Metropolis Hastings algorithm for sampling from a BayesEL posterior.  Under our setup, (see Section \ref{section-max}) the parameters of interest are split into two sets.  New values of the first set of parameters are proposed first.  
Next, by using the estimating equations and weights obtained from a first-stage empirical likelihood, we find a trial value of the remaining parameters.  The new values of the remaining parameters are then proposed depending on this trial value. 
The two-step procedure ensures that for a proposed value of the first set of parameters in their marginal support, the remaining parameters would be in their marginal support with a hiogh probability.  This procedure is not affected by the non-convexity of the support and improves the acceptance rate of the resulting chain. 
Our method does not require the estimating equation to be smooth, even continuous, in the parameters.  We extend it to a reversible jump Markov chain Monte Carlo scheme which allows us to successfully implement empirical likelihood in Bayesian model selection. 



\section{A maximum conditional empirical likelihood estimator with Applications to Bayesian Empirical Likelihood Computation}\label{section-max}

\subsection{Setup of the Bayesian Empirical Likelihood}



Consider independent observations $x_1,\ldots, x_n$ from an unknown distribution $F^0$, which depends on a parameter vector $\theta=(\theta_1,\theta_2)\in \Theta=\Theta_1\times\Theta_2$ of length $p+q$. Assume that the prior knowledge on  $(\theta_1,\theta_2)$ is specified by $\pi(\theta_1,\theta_2)$.  
Suppose the unknown distribution $F^0$ satisfies certain estimating equations.  Some of these equations do not depend on $\theta_2$, while the others are functions of $\theta_2$ and possibly of all parameters. More specifically, let

\begin{equation}\label{estimating equation}
E_{F^0}[g(x,\theta^0_1)]=0 \quad\text{and}\quad E_{F^0}[h(x,\theta^0_1,\theta^0_2)]=0,
\end{equation}
where $g(x,\theta_1)\in \mathbb{R}^l$ and $h(x,\theta_1,\theta_2)\in\mathbb{R}^d$ and $\theta^0$ is the true value of $\theta$.

Estimating equations with such structures appear in traditional linear models, graphical Markov models \citep{lauritzen_book_1996}, and models with patterned missing data eg. two-phase designs, models with surrogate variables \citep{qin_zhang_leung_2009} etc.  

Let $\mathcal{F}_{\theta}$ be a set of distribution functions depending on parameter $\theta\in\Theta$.  for any $F\in \mathcal{F}_{\theta}$ its empirical likelihood is derived by maximising the so called ``non-parametric likelihood" 

\begin{equation}\label{nonparametric}
\mc{L}(F)=\prod_{i=1}^n \{F(x_i)-\mathop{Lim}_{h\downarrow0}F(x_i-h)\}=\prod_{i=1}^n \{F(x_i)-F(x_i-)\},
\end{equation}
over $\mathcal{F}_\theta$ under constraints, depending on $g(x,\theta_1)$ and $h(x,\theta_1,\theta_2)$.

Suppose, $\omega_i=F(x_i)-F(x_i-)$ and $\omega=(\omega_1,\ldots,\omega_n)$ is the vector of weights on the components of $x=(x_1$, $\ldots$, $x_n)$. Given $\theta_1\in\Theta_1$ and $ \theta_2\in\Theta_2$, we define a constrained set of $\omega$, which depends on $g(x,\theta_1)$ and $h(x,\theta_1,\theta_2)$ as 

\begin{equation}\label{W_G}
\mathcal{W}_{G}(\theta_1)=\left\{\omega: \sum_{i=1}^n \omega_ig(x_i,\theta_1)=0\right\}\cap\Delta_{n-1},
\end{equation}

\begin{equation}\label{W_H}
\mathcal{W}_{H}(\theta_1,\theta_2)=\left\{\omega: \sum_{i=1}^n \omega_ih(x_i,\theta_1,\theta_2)=0\right\}\cap\Delta_{n-1},
\end{equation}

and

\begin{equation}\label{W}
\mathcal{W}(\theta_1,\theta_2)=\mathcal{W}_{G}(\theta_1)\cap\mathcal{W}_{H}(\theta_1,\theta_2).
\end{equation}

The empirical likelihood of $F$, which can also be expressed as a likelihood of $\theta$ is given by

\begin{equation}\label{el}
\mc{L}(\theta_1,\theta_2)=\mc{L}(F)=\max_{\omega\in\mathcal{W}(\theta_1,\theta_2)}\prod_{i=1}^n \omega_i(\theta_1,\theta_2).
\end{equation}
  
\noindent We define, $\mc{L}(\theta_1,\theta_2)=\mc{L}(F)=0$, when the problem \eqref{el} is infeasible.


Now by using $\mc{L}(\theta_1,\theta_2)$ as a likelihood, one defines Bayesian empirical likelihood (BayesEL) posterior as
\begin{equation}\label{posterior}
\Pi_{EL}(\theta_1,\theta_2\mid x)=\frac{ \mc{L}(\theta_1,\theta_2)\pi(\theta_1,\theta_2)}{\int \mc{L}(\theta_1,\theta_2)\pi(\theta_1,\theta_2)d\theta_1d\theta_2}.
\end{equation}


The empirical likelihood is based on an empirical estimate of $F_0$ which satisfies the estimating equations in \eqref{estimating equation}.  Clearly, for any continuous $F$, the likelihood $\mc{L}(F)=0$. Thus, by construction this estimate is discrete.  
Furthermore, by definition, $\omega\in\Delta_{n-1}$, that is $\mc{L}(\theta_1,\theta_2)$ is bounded by one.  This implies the posterior $\Pi_{EL}$ is proper for any proper prior.


Validity of empirical and similar such likelihoods in 
Bayesian inference has been a topic with extensive discussion. Using a criterion  proposed by \citet{monahan1992proper}, \citet{lazar2003bayesian} examined the validity of BayesEL procedures by Monte Carlo simulations.
Asymptotic frequentist coverages of Bayesian credible sets for a general empirical likelihood formulation was studied by \citet{fang2006empirical}, \citet{corcoran_1998}.  
Higher order asymptotics of BayesEL posterior have been studied by \citet{zhong_ghosh_2017,sueishi_2024}.  

By formulation, a BayesEL posterior only needs a specification of the model in terms of estimating equations which are unbiased under the truth.  Thus, by using BayesEL procedure, one can avoid making non-testable assumptions about the data distribution.  
This has been found to be useful and the BayesEL procedures have seen many applications on different problems in recent times.   Examples include complex surveys \citep{rao2010bayesian,zhaoGhoshRaoWu2020},  small area estimation \citep{chaudhuri2011empirical,porter_holan_wickle_2015}, quantile regression \citep{yang2012bayesian} among others. 


Becuase of absence of an analytic form, any BayesEL inference requires samples drawn from the resulting posterior using a suitable Markov chain Monte Carlo (MCMC) sampler.  For BayesEL, however, efficient adaptive proposals for MCMC sampling are not easily constructed, and has been a big bottleneck to its applicability. This is primarily due to the complex nature of the support of the empirical likelihood. 

Provided the prior is supported over whole $\Theta$,  the BayesEL posterior is positive,  if the maximisation problem in \eqref{el} is feasible for the particular $\theta_1$ and $\theta_2$.  
This happens when $\hat{\omega}_i(\theta_1,\theta_2)> 0$, for all $i=1$, $\ldots$, $n$, which in turn happens if and only if the origin is in the convex hull of the points $\left(g(x_i,\theta_1), h(x_i,\theta_1,\theta_2)\right)$, $i=1$, $\ldots$, $n$. 
However, as seen in Example \ref{exm:muVar} below, usually, this condition is not satisfied over the whole $\Theta$.  For certain values of $\theta\in \theta$, the problem in \eqref{el} is infeasible and the corresponding set $\mathcal{W}(\theta_1,\theta_2)$ would be empty.

\begin{exm}\label{exm:muVar}
  Consider $10$ independently and identically distributed univariate observations $x_i, i=1,\ldots,n = 10$, generated from a normal distribution with mean $\mu = 0$ and variance $\sigma^2 = 1$.  
Our goal here is to estimate $\mu$ and $\sigma^2$ from the data.
 
Suppose we assume diffused but proper prior distributions $\pi(\mu)$ a $\mathcal{N}(0,100)$ and $\pi(\sigma^2)$
an Inverse Gamma $\mathcal{IG}(.001,.001)$ for $\mu$ and $\sigma^2$ respectively.  The full parametric joint posterior distribution of $(\mu, \sigma^2)$ is given by,
\begin{equation}\label{eq:normpost}
\Pi_{N}(\mu,\sigma^2|x) = \frac{(2\pi\sigma^2)^{-n/2}\exp\left\{-(2\sigma^2)^{-1}\sum^n_{i=1}(x_i-\mu)^2\right\}\pi(\mu)\pi(\sigma^2)}{\int (2\pi\sigma^2)^{-n/2}\exp\left\{-(2\sigma^2)^{-1}\sum^n_{i=1}(x_i-\mu)^2\right\}\pi(\mu)\pi(\sigma^2) d\mu d\sigma^2}.
\end{equation}

For the BayesEL posterior, using the definition of mean and variance, it is clear that,
\begin{equation}\label{eqn1}
E_{F^0}\left[x_i-\mu_0\right] = 0 ~~~~\text{and}~~~~E_{F^0}\left[(x_i-\mu_0)^2-\sigma^2_0\right] = 0.
\end{equation}

\begin{figure}[t]
\begin{center}
 \subfigure[\label{fig1subfig1}]{
  \resizebox{.45\columnwidth}{!}{\includegraphics{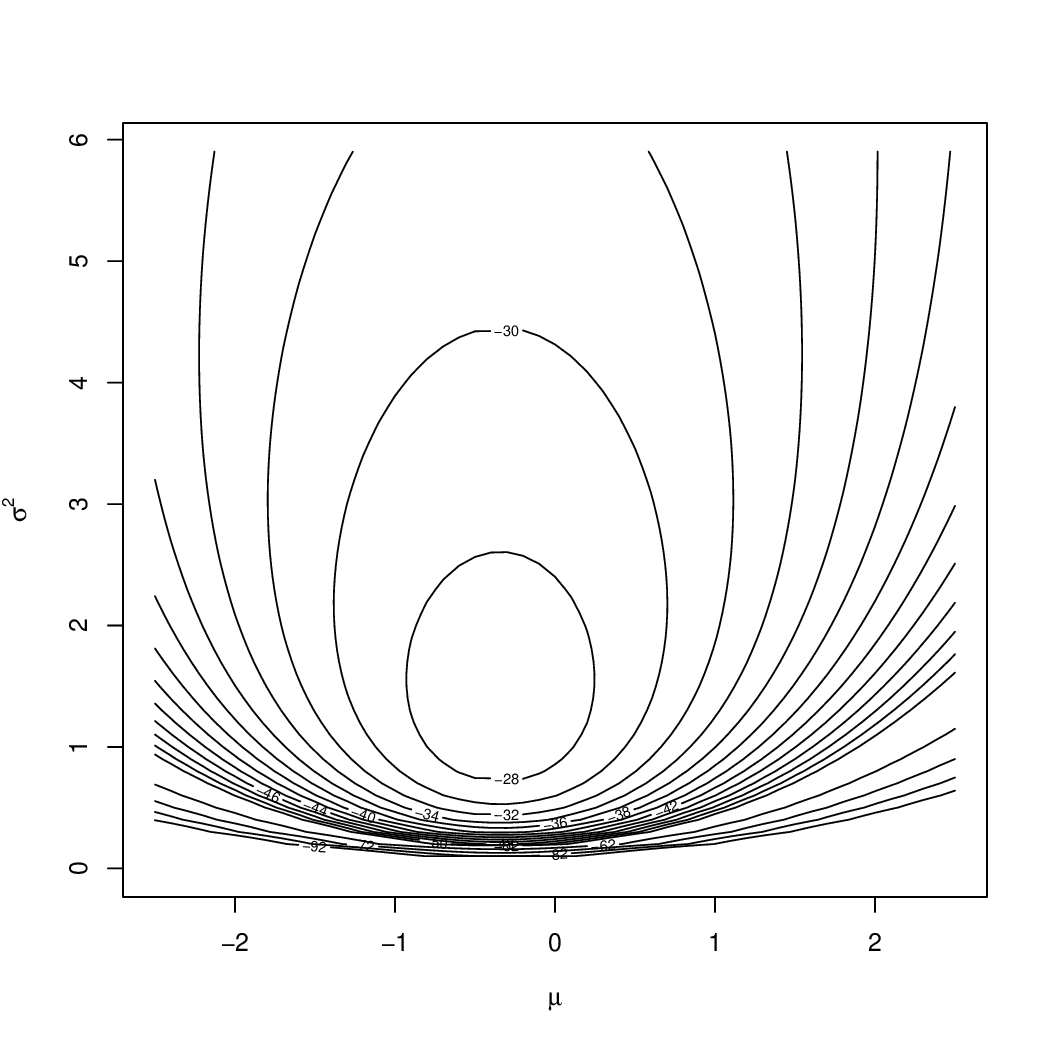}}
   }
 \subfigure[ \label{fig1subfig2}]{
  \resizebox{.45\columnwidth}{!}{\includegraphics{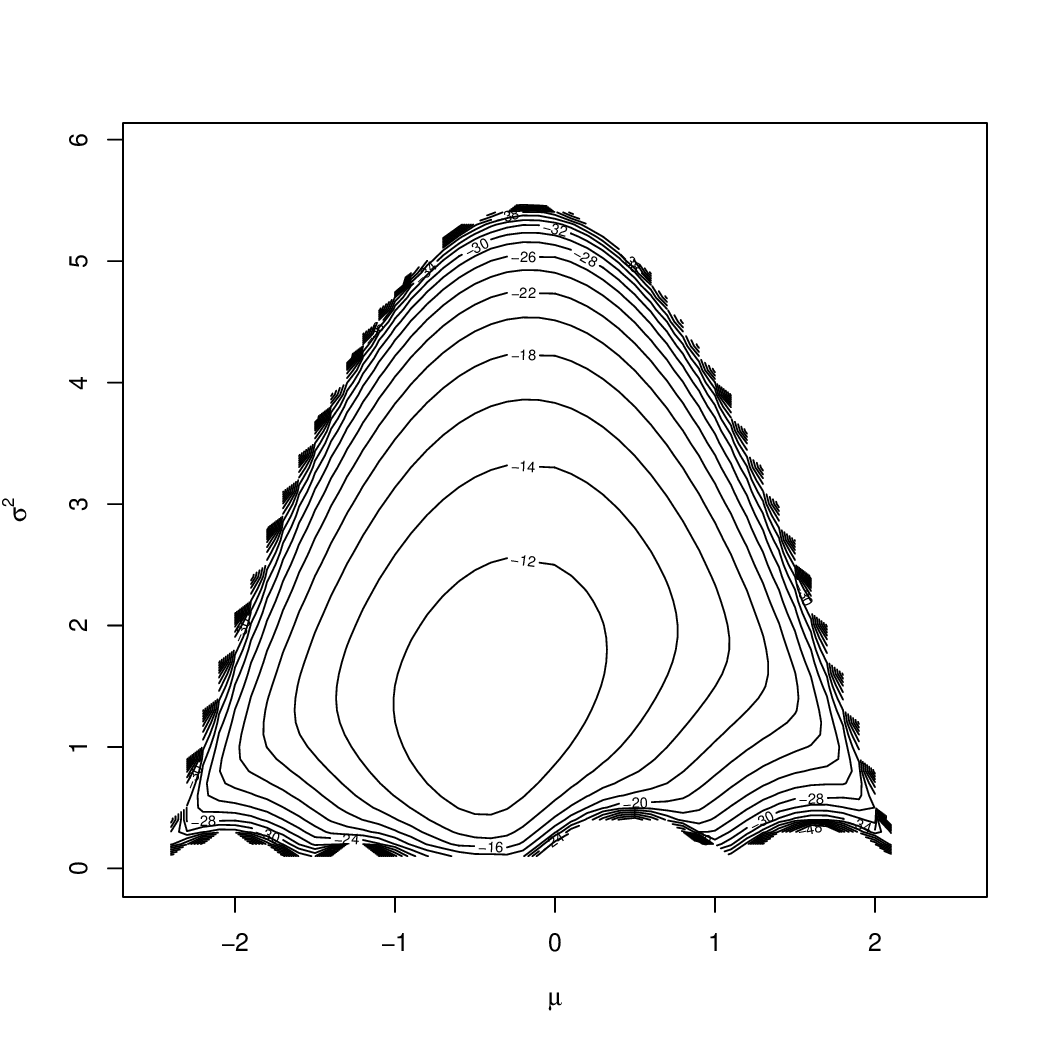}}
   }
\end{center}
\caption{The contour plots of the logarithms of the numerators of (a) \eqref{eq:normpost} ie. $\log \Pi_{N}(\mu,\sigma^2|x)$ and (b) \eqref{posterior} ie. $\log \Pi_{EL}(\mu,\sigma^2|x)$ for different values of $\mu$ and $\sigma^2$.} 
\label{fig1}

\end{figure}

In Figure \ref{fig1}, we present the contour plots of the logarithm of the numerators of the expressions in \eqref{eq:normpost} and \eqref{posterior} for different values of $\mu$ and $\sigma^2$.  Our motivation here is to compare the supports of 
$\Pi_{N}(\mu,\sigma^2|x)$ and $\Pi_{EL}(\mu,\sigma^2|x)$.  Since the normal distribution is supported over the whole real line, $\Pi_{N}$ is supported over the whole half plane.  
However, for many values of $\mu$ and $\sigma^2$, the problem in \eqref{el} may be infeasible.  Thus for those values by definition, $\mc{L}(\mu,\sigma^2)=0$.  That is, $\Pi_{EL}$ may not be supported on the whole half plane.\hfill\qed

\end{exm}

Even though numerous authors have discussed such \emph{empty-set problems} and its remedies in the frequentist paradigm \citep{chen2008adjusted,emerson2009calibration,liu2010adjusted, tsao2013empirical,tsao2013extending},  relatively little is known about the properties of empirical likelihood support.  Under certain conditions \citet{chaudhuri_mondal_yin_2017} show that the support is an open set.  
Even for simple models though, the posterior support would be non-convex.  The marginal support of $\theta_2$ may depend very much on the values of $\theta_1$ and may vary dramatically with new proposed value of the latter. In such cases, usual random walk proposals will not be efficient.  Some adaptive MCMC procedure would be required.  

The random walk based adaptive procedures e.g \citet{haario2001adaptive} do not take into account the non-convexity of the support and as a result is not very efficient.  Methods like parallel tempering \citep{geyer1992markov, liu2008monte} too mixes very slowly.     

One adaptive procedure have been discussed by \citet{chaudhuri_mondal_yin_2017}, who show that the gradient of the log-empirical likelihood diverges at the boundary of support and use this information to design an Hamiltonian Monte Carlo procedure to sample from a BayesEL posterior.  
The diverging derivative ensures that when the chain approaches the boundary it reflects back towards the centre of the support and almost never steps out. 

Even though the HMC procedure is useful in many problems, it requires the estimating equations to be smooth in terms of the parameters.  In many applications this condition does not hold.  In fact, in many cases (eg. estimating quantiles) the estimating equations may not even be continuous function of the parameters.

We introduce a basic two-step Metropolis Hastings procedure to sample from BayesEL posteriors below. This method does not put any smoothness conditions on the estimating equations and thus can be applied to all the situations described above.  
Furthermore, we develop a BayesEL model selection procedure for which we propose a RJMCMC procedure which provides the only known efficient way to sample from the resulting posterior.

\subsection{A Two-step Maximum Conditional Empirical Likelihood Estimator} 
Under our setup, efficient moves for Metropolis Hastings procedure can be achieved if for a given value of $\theta_1\in\Theta_1$ we can propose a value of $\theta_2$ such that the empirical likelihood $\mc{L}(\theta_1,\theta_2)$ is relatively large.  This can be ensured by judicious use of the estimating equations in \eqref{W_G} and \eqref{W_H} above.  

Suppose $\theta_1=a\in\Theta_1$ is fixed.  We define the \emph{maximum conditional empirical likelihood estimator} (MCELE) of $\theta_2$ as
\begin{equation}
\Tet(a)=\argmax_{\theta_2\in\Theta_2}\prod_{i=1}^n\hat\omega_i(a,\theta_2)=\argmax_{\theta_2\in\Theta_2, \omega\in\mathcal{W}(a,\theta_2) }\left\{\prod_{i=1}^n \omega_i\right\}.\label{max cond theta_2 2}
\end{equation}

Clearly, by definition, for a given $\theta_1=a$, $\mc{L}(a,\Tet(a))$ is the highest possible value of the empirical likelihood.  We use $\Tet$ in our proposed procedure. 

There are several ways to compute $\Tet(a)$. Equation \eqref{max cond theta_2 2} indicates that it can be obtained from a two-stage maximisation. 
 Our setup allows an alternative characterisation of $\Tet(a)$ which reduces the cost of its computation. This characterisation is also key to the two-step Markov chain Monte Carlo method for drawing sample from the BayesEL posterior in \eqref{posterior}.

With the definition of $\mathcal{W}_G(a)$ is \eqref{W_G}, suppose we define
\begin{equation}\label{trial problem}
\hat\nu(a)=\argmax_{\nu\in\mathcal{W}_{G}(a)}\sum_{i=1}^n \log\nu_i.
\end{equation}

By substituting $\hat\nu(a)$ for $\omega$ in \eqref{W_H}, we get the following equations in $\theta_2$
\begin{equation}\label{eqn of theta_2}
\sum_{i=1}^n\hat{\nu}_i(a)h(x_i,a,\theta_2)=0.
\end{equation}

It is easily seen that the following result holds.
\begin{theorem}\label{chapter2maxcond}
 Suppose $\tilde\theta_2$ solves   \eqref{eqn of theta_2}. Then $\tilde\theta_2=\Tet(a)$.
\end{theorem}

Our proposed maximum conditional estimator of $\theta_2$ given $\theta_1=a$ is motivated partly by \cite{chaudhuri2008generalized}, who used similar procedure in a frequentist setting. We shall see later that computationally, the characterisation in Theorem \ref{chapter2maxcond} is very convenient. 
For a known value $a$, maximising the product of weights over $\mathcal{W}_G(a)$ is a convex problem. Thus, it would have an unique solution.  Thereafter, given the optimal weights $\hat\nu$, 
one has to solve \eqref{eqn of theta_2} at the worst numerically, which is often easy. In many cases, analytic solutions can be found.  The solution $\tilde\theta_2$ form the basis of the two-step MCMC procedure we describe next.





\section{A two-step Metropolis Hastings method for fixed dimensional state space}\label{section fixed}

We now describe the two-step Metropolis Hastings method to sample from the BayesEL posterior when the dimension of the parameter $\theta\in\Theta$ remains fixed.  
We shall assume the setup in Section \ref{section-max} and for any value of $\theta_1\in\Theta_1$ such that \eqref{trial problem} is feasible, there is an unique $\Tet$ which solves \eqref{eqn of theta_2}.

Our proposed two-step method is based on the following intuition. First of all notice that if $\theta_1=a$, from \eqref{trial problem} and \eqref{eqn of theta_2} we get,
\[
\mc{L}\left(a,\Tet(a)\right)=\prod^n_{i=1}\hat{\nu}_i(a),
\]

provided, the problem in \eqref{trial problem} is feasible.  For any $\theta_2$ sufficiently close to $\Tet(a)$, by continuity of the weights and the likelihood, $L\left(a,\theta_2\right)$ would be large and thus in a Metropolis Hastings sampler, the move to the proposed point $(a,\theta_2)$ would have a higher probability of getting accepted. 

 More specifically (see Algorithm \ref{algorithm1}), suppose at iteration $t$, the chain is at $\theta^{(t)}=(\theta^{(t)}_1,\theta^{(t)}_2)$.  We first propose a value of $\theta_1$, 
denoted as $\theta_1^{(t+1)}$ from a proposal distribution $q_1$ 
possibly depending on $\theta_1^{(t)}$.  In the second step, we compute the corresponding MCELE $\Tet\left(\theta_1^{(t+1)}\right)$.  

A new value of $\theta_2$, denoted by $\theta_2^{(t+1)}$, is then proposed from a proposal distribution $q_2$, 
possibly depending on $\theta_1^{(t+1)}$ and $\Tet\left(\theta_1^{(t+1)}\right)$.  
Finally, similar to usual the Metropolis Hastings algorithm, the proposed point $(\theta_1^{(t+1)}, \theta_2^{(t+1 )})$ is accepted with probability in \eqref{ln:s} of Algorithm \ref{algorithm1}.

\begin{algorithm}[t]
\caption{The two-step Metropolis Hastings algorithm for fixed dimension}
\label{algorithm1}
\begin{algorithmic}[2]
\REQUIRE $\theta^{(1)}=\left(\theta_1^{(1)},\theta_2^{(1)}\right)$.

\FOR{$t=1$ to $L$}
\STATE Propose $\theta_1^{(t+1)}$ following density $q_1(\cdot\mid \theta_1^{(t)})$;
\IF{ the problem \eqref{trial problem} is infeasible for $\theta_1^{(t+1)}$}
\STATE $\theta^{(t+1)}\leftarrow\theta^{(t)}$;
\ELSE
\STATE Compute $\hat\nu\left(\theta_1^{(t+1)}\right)$ by solving problem in \eqref{trial problem};
\STATE Compute $\tilde{\theta}_2^{(t+1)}=\Tet\left(\theta_1^{(t+1)}\right)$ by solving equation \eqref{eqn of theta_2} with weights $\hat\nu\left(\theta_1^{(t+1)}\right)$;
\STATE Propose ${\theta_2}^{(t+1)}$  following  density $q_2\left(\cdot\mid\theta_1^{(t+1)},\tilde{\theta}_2^{(t+1)}\right)$; 
\STATE Calculate the empirical likelihood $\mc{L}(\theta_1^{(t+1)},\theta_2^{(t+1)})$;
\STATE Accept $\theta^{(t+1)}=(\theta_1^{(t+1)},\theta_2^{(t+1)})$ with probability $\alpha$, where
{\small
\begin{equation}\label{ln:s}
\alpha=\min\left\{1, \frac{\mc{L}(\theta_1^{(t+1)},\theta_2^{(t+1)})\pi(\theta_1^{(t+1)},\theta_2^{(t+1)})q_2(\theta_2^{(t)}|\theta_1^{(t)},\tilde{\theta}_2^{(t)})q_1(\theta_1^{(t)}\mid \theta_1^{(t+1)})}{\mc{L}(\theta_1^{(t)},\theta_2^{(t)})\pi(\theta_1^{(t)},\theta_2^{(t)})q_2(\theta_2^{(t+1)}|\theta_1^{(t+1)},\tilde{\theta}_2^{(t+1)})q_1(\theta_1^{(t+1)}\mid \theta_1^{(t)})}\right\};
\end{equation}
}
\ENDIF
\ENDFOR
\RETURN $(\theta_1,\theta_2)^{(1:L)}$.
\end{algorithmic}
\end{algorithm}
 
 Suppose $L$ is the length of the sequence and for any $t$, we denote $(\theta_1,\theta_2)^{(1:t)} = \left(\theta_1^{(1:t)},\theta_2^{(1:t)}\right)=\left\{(\theta_1^{(1)},\theta_2^{(1)}),\ldots,(\theta_1^{(t)},\theta_2^{(t)})\right\}$ 
to be the sequence of observations up to time $t$ obtained from Algorithm \ref{algorithm1} from the beginning (see also Figure \ref{THMfigure}).  By construction, the sequence $(\theta_1,\theta_2)^{(1:L)}$ has following properties. 

 \begin{enumerate}
 \item $\theta_1^{(1:L)}$ is a Markov chain, i.e. for each $t=1,2,\ldots,T-1$, $\theta_1^{(t+1)}\ci \theta_1^{(1:t-1)} \big|\theta_1^{(t)}$.
 \item Given $\theta_1^{(t)}$, $\theta_1^{(t+1)}$  is independent of $\theta_2^{(1:t)}$, i.e. $\theta_1^{(t+1)}\ci \theta_2^{(1:t)}\big|\theta_1^{(t)}$.
\item For all $t$, $\Tet\left(\theta_1^{(t)}\right)$ is a deterministic function of $\theta_1^{(t)}$.
 \item $\theta_2^{(t+1)}$ is independent of $\theta^{(1:t)}$ given $\theta_1^{(t+1)}$, i.e. $\theta_2^{(t+1)}\ci \left(\theta_1^{(1:t)},\theta_2^{(1:t)}\right)\big|\theta_1^{(t+1)}$.
\end{enumerate}

By construction, $(\theta_1^{(1:L)},\theta_2^{(1:L)})$ is jointly a Markov chain and its transition probability can be determined. 

\begin{theorem}\label{chapter3thm1}
Suppose the sequence $(\theta_1,\theta_2)^{(1:L)}$ is generated from the Algorithm \ref{algorithm1}. Assume that for each $\theta_1^{(t+1)}$, $\Tet\left(\theta_1^{(t+1)}\right)$  
is the unique solution of the equation \eqref{eqn of theta_2}.  Then $(\theta_1,\theta_2)^{(1:L)}$ is a Markov chain and its transition probability is given by
\begin{equation}\label{tran}
P\left(\theta_1^{(t+1)},\theta_2^{(t+1)}\mid\theta_1^{(t)},\theta_2^{(t)}\right)=P\left(\theta_2^{(t+1)}\mid \theta_1^{(t+1)},\Tet\left(\theta_1^{(t+1)}\right)\right)P\left(\theta_1^{(t+1)}\mid\theta_1^{(t)}\right).
\end{equation}
\end{theorem}

\begin{figure}[t]
\begin{center}
\resizebox{.75\columnwidth}{!}{\input{figure1_paper.tex}}
\end{center}
\medskip
\caption{A schematic diagram of Algorithm \ref{algorithm1}.  The dashed lines indicate fully deterministic relationship, where as the solid lines show stochastic relationship.  The variables in the dotted box are proposed at each step.}
\label{THMfigure}
\end{figure}

Theorem \ref{chapter3thm1} shows that jointly $\left(\theta_1,\theta_2\right)^{(1:L)}$ is a Markov chain and the transition probability is the product of $q_1$
and $q_2$
However, marginally, $\theta_2^{(1:L)}$ itself is not a Markov chain (see Figure \ref{THMfigure}).  
The chain $(\theta_1,\theta_2)^{(1:L)}$ is reversible because the chain $\theta_1^{(1:L)}$ is. 
The irreducibility and aperiodicity of the chain can be ensured by judicious choice of the proposal distributions $q_1$ and $q_2$.  A choice such that the support of 
the joint proposal $q_1\otimes q_2$ covers the whole of $\Theta_1\times\Theta_2$ would generally suffice.  
  Under such choices, the chain will converge to its stationary distribution $\Pi_{EL}(\theta_1,\theta_2\mid x)$.

In practical implementation of the two step method, $\theta_1$ can be proposed in many ways.  One can use traditional variations of random walk proposals or, provided the estimating equations are smooth, Hamiltonian Monte Carlo \citep{chaudhuri_mondal_yin_2017} or similar methods.  
The proposal distribution in the second step has to be independent of the previous steps and depend on the MCELE. In fact, in most cases, $q_2$ can be chosen only to depend on the MCELE $\Tet(\theta_1^{(t+1)})$.

\section{BayesEL Model Selection} 

We now turn to Bayesian variable selection using empirical likelihood.  Such problems can now be attempted because our two-step Metropolis Hastings algorithm can be extended to a reversible jump Markov chain Monte Carlo (RJMCMC) procedure, which can efficiently draw samples from the posterior over discrete model space.  
\subsection{Setup}
We first discuss the constraints under which the empirical likelihood is computed.  We will concentrate on the linear models, however, the formulation would easily extend to other models as well.
 
Suppose the response variable is denoted by $y$ and there are $s$ potential covariates, where $s$ is assumed to be strictly smaller than the sample size $n$.  
A model is specified by a binary vector $\pmb{\gamma}=(\gamma_1,\ldots,\gamma_s)$, where $\gamma_i$ is $1$ if the $i$th covariate is included in the model and $0$ otherwise.  
For a given $\pmb{\gamma}$, $\pmb{x_\gamma}=(x_{\star j} : \gamma_j=1, j\in\{1,\ldots,s\})$, a subset of columns of $n\times s$ data matrix $x$, is the matrix of covariates under model $\pmb{\gamma}$.  
The corresponding coefficient vector is denoted by $\pmb{\beta_\gamma}=(\beta_{i} : \gamma_i=1, i\in\{1,\ldots,s\})$. 

For a given model $\pmb{\gamma}$, the linear model can be be described by,

\begin{equation}\label{linear}
E\left[y\mid \pmb{x}_{\pmb{\gamma}}\right]=\pmb{x}_{\pmb{\gamma}}\pmb{\beta_{\gamma}}\quad\text{and}\quad\text{Var}\left[y\mid \pmb{x}_{\pmb{\gamma}}\right]=\sigma^{2}_{\pmb{\gamma}},
\end{equation}
where, the expectations are taken with respect to the unknown true distribution.

In order to define the BayesEL posterior, as before, we assign unknown weights $\omega_i$ to observation $x_i$.  For a model $\pmb{\gamma}$, parameters $\pmb{\beta_{\gamma}}$ and $\sigma^{2}_{\pmb{\gamma}}$, the empirical likelihood is defined as:
 
\begin{equation}\label{eq:elms}
\mc{L}(\pmb{\gamma},\pmb{\beta_{\gamma}},\sigma^{2}_{\pmb{\gamma}})=\max_{\omega\in\mathcal{W}\left(\pmb{\gamma},\pmb{\beta}_{\pmb{\gamma}},\sigma^2_{\pmb{\gamma}}\right)}\prod^n_{i=1}\omega_i,
\end{equation} 
where 

\begin{align}
\mathcal{W}_{0}\left(\pmb{\beta}_{\pmb{\gamma}}\right)&=\left\{\omega~:~\omega^T\left(y-\pmb{x}_{\pmb{\gamma}}\pmb{\beta_{\gamma}}\right)=0\right\}\cap\Delta_{n-1},\label{eq:int}\\
\mathcal{W}_{\gamma}\left(\pmb{\beta}_{\pmb{\gamma}}\right)&=\bigcap_{j:\gamma_j=1}\left\{\omega~:~x^T_{\star j}D_{\omega}\left(y-\pmb{x}_{\pmb{\gamma}}\pmb{\beta_{\gamma}}\right)=0\right\}\cap\Delta_{n-1},\label{eq:model}\\
\mathcal{W}_{\gamma^c}\left(\pmb{\beta}_{\pmb{\gamma}}\right)&=\bigcap_{j:\gamma_j=0}\left\{\omega~:~x^T_{\star j}D_{\omega}\left(y-\pmb{x}_{\pmb{\gamma}}\pmb{\beta_{\gamma}}\right)=0\right\}\cap\Delta_{n-1},\label{eq:modelc}\\
\mathcal{W}_{\sigma}\left(\pmb{\beta}_{\pmb{\gamma}},\sigma^2_{\pmb{\gamma}}\right)&=\left\{\omega~:~\left(y-\pmb{x}_{\pmb{\gamma}}\pmb{\beta_{\gamma}}\right)^TD_{\omega}\left(y-\pmb{x}^T_{\pmb{\gamma}}\pmb{\beta_{\gamma}}\right)-\sigma^{2}_{\pmb{\gamma}}=0\right\}\cap\Delta_{n-1},\label{eq:sigma}\\
\mathcal{W}\left(\pmb{\gamma},\pmb{\beta}_{\pmb{\gamma}},\sigma^2_{\pmb{\gamma}}\right)&=\mathcal{W}_{0}\left(\pmb{\beta}_{\pmb{\gamma}}\right)\cap\mathcal{W}_{\gamma}\left(\pmb{\beta}_{\pmb{\gamma}}\right)\cap\mathcal{W}_{\gamma^c}\left(\pmb{\beta}_{\pmb{\gamma}}\right)\cap\mathcal{W}_{\sigma}\left(\pmb{\beta}_{\pmb{\gamma}},\sigma^2_{\pmb{\gamma}}\right).\label{eq:wms}
\end{align}
Here $D_{\omega}$ is the $n\times n$ diagonal matrix with $\omega$ as the diagonal.  As before, $\mc{L}(\pmb{\gamma},\pmb{\beta_{\gamma}},\sigma^{2}_{\pmb{\gamma}})=0$ if the problem in \eqref{eq:elms} is infeasible.


The set of constraints described above goes beyond the model specification in \eqref{linear}.  It is readily seen that $\mathcal{W}_{0}$ and $\mathcal{W}_{\sigma}$ ensure that the expectation and variance of the residuals under estimated empirical distribution are zero and $\sigma^{2}_{\pmb{\gamma}}$ respectively.  
The set $\mathcal{W}_{\gamma}$ implies that the residuals from model $\pmb{\gamma}$ are uncorrelated to the covariates in the model.  These constraints follow from the score equations of the model.  
The constraints in $\mathcal{W}_{\gamma^c}$ demand some explanation.  Here we impose that the residual from model $\pmb{\gamma}$  is uncorrelated to all the available variables absent from the model.  This constraints do not follow directly from the linear model in \eqref{linear}.  However, these constraints can be justified from a predictive modelling consideration.  
Clearly, if a covariate not in the current model is correlated to the residuals, its inclusion in the model is likely to improve prediction. 
Several authors (eg. \citet{variyath2010empirical}, \citet{kolaczyk1995information}, \citet{chenWangWuLi2022}) have used the same setup in context of frequentist model selection in generalised linear and moment condition models.   

Once the likelihood $\mc{L}(\pmb{\gamma},\pmb{\beta_{\gamma}},\sigma^{2}_{\pmb{\gamma}})$, the priors $\pi(\pmb{\gamma})$ and $\pi(\pmb{\beta_{\gamma}},\sigma^{2}_{\pmb{\gamma}})$ has been determined the BayesEL posterior can be defined as:
\begin{align}
\Pi_{EL}\left(\pmb{\gamma},\pmb{\beta_{\gamma}},\sigma^{2}_{\pmb{\gamma}}\mid y,x\right)&=\frac{\mc{L}(\pmb{\gamma},\pmb{\beta_{\gamma}},\sigma^{2}_{\pmb{\gamma}})\pi(\pmb{\beta_{\gamma}},\sigma^{2}_{\pmb{\gamma}})\pi(\pmb{\gamma})}{\int \mc{L}(\pmb{\gamma},\pmb{\beta_{\gamma}},\sigma^{2}_{\pmb{\gamma}})\pi(\pmb{\beta_{\gamma}},\sigma^{2}_{\pmb{\gamma}})\pi(\pmb{\gamma})d\pmb{\gamma}d\pmb{\beta_{\gamma}}d\sigma^{2}_{\pmb{\gamma}}}\nonumber\\
&\propto \left\{\max_{\omega\in\mathcal{W}\left(\pmb{\gamma},\pmb{\beta}_{\pmb{\gamma}},\sigma^2_{\pmb{\gamma}}\right)}\prod^n_{i=1}\omega_i\right\}\pi(\pmb{\beta_{\gamma}},\sigma^{2}_{\pmb{\gamma}})\pi(\pmb{\gamma}).\label{eq:postms}
\end{align}


Sampling from a BayesEL posterior is difficult when the dimension of the parameter space can vary between iterations.  In fully parametric setups reversible jump Markov chain Monte Carlo (RJMCMC) samplers \citep{green1995reversible} are generally used in similar situations.  Many efficient RJMCMC procedures have been studied (see \citet{fan2010reversible},\citet{dellaportas2002bayesian},\citet{robert2002bayesian}) in such settings.  
For BayesEL model selection sampling from the posterior in \eqref{eq:postms} could be more challenging.  Under different models, the posterior supports could be quite different.  Proposing parameter values in the new marginal supports specially in a cross-model move would not be easy.  
The HMC procedure in \citet{chaudhuri_mondal_yin_2017} cannot be applied here. 
The model space is discrete and none of the estimating equations could be a smooth functions of the models as parameters.  However, our two-step Metropolis Hastings algorithm can be extended to an efficient reversible jump Markov chain Monte Carlo procedure.

\subsection{Two-Step Reversible Jump Markov Chain Monte Carlo for BayesEL Linear Model Selection}
 A reversible jump Markov chain Monte Carlo (RJMCMC) algorithm has two stages. First, it updates parameters within a model.  Second, it updates parameters from one model to another. 
 This stage requires a cross-model proposal that can propose values, which are more likely to be accepted, in the new model.  The main problem is therefore to construct such cross-model proposal on which the efficiency of the RJMCMC algorithm depends. 

\begin{figure}[t]
\input{figure2_bare.tex}
\caption{A schematic diagram of the proposed RJMCMC Algorithm \ref{algorithm2}. Dashed arrows indicate fully deterministic relationship.  The variables in dotted 
boxes are proposed in the first stage where the parameters within the model is updated.  
The variables in the dashed boxes are proposed in the second stage to update the model.}
\label{fig:rjmcmc}
\end{figure}

Suppose the dimension of model is defined by $\sum^n_{i=1}\gamma_i$.  Let the current model be $(\pmb{\gamma},\pmb{\beta_\gamma},\sigma^2_{\pmb{\gamma}})$ with dimension $k$ and we want to propose a jump to a model $(\pmb{\gamma^{\prime}},\pmb{\beta_{\gamma^{\prime}}},\sigma^2_{\pmb{\gamma^{\prime}}})$ 
with dimension $k^{\prime}=k+1$, where the binary vector $\pmb{\gamma^{\prime}}-\pmb{\gamma}$ has $0$ at all components except the $j$th one (i.e. the $j$th covariate is added to the model). The cross-model proposal is constructed in two steps. 
First,  for the purpose of dimensional matching, a random variable $u$ is generated from a proposal distribution $q_U$.  This ensures $dim(\pmb{\beta_{\gamma}})+dim(u)=dim(\pmb{\beta_{\gamma'}})$.  Second, we define the map 

\begin{equation}
(\pmb{\beta}_{\pmb{\gamma}^{\prime},-j}, \pmb{\beta}_{\pmb{\gamma}^{\prime},j}, \sigma^2_{\pmb{\gamma^{\prime}}})=g_{\gamma\to\gamma^{\prime}}(\pmb{\beta_\gamma},u, \sigma^2_{\pmb{\gamma}})=\left(\pmb{\beta_{\gamma}}+(\hat{\pmb{\beta}}_{\pmb{\gamma}',-j}-\hat{\pmb{\beta}}_{\pmb{\gamma}}),u+\hat{\beta}_{\pmb{\gamma^{\prime}},j},\sigma^2_{\pmb{\gamma}}+(\hat\sigma^2_{\pmb{\gamma^{\prime}}}-\hat\sigma^2_{\pmb{\gamma}})\right).\label{eq:g}
\end{equation}

Here, $\hat{\pmb{\beta}}_{\pmb{\gamma}}$ and $\hat{\beta}_{\pmb\gamma^{\prime}}$ are the ordinary least squares estimates of the regression coefficients for the model $\pmb{\gamma}$ and $\pmb{\gamma}^{\prime}$ respectively.  By $\pmb{\beta}_{\pmb{\gamma},j}$ we denote the $j$th component of the 
vector $\pmb{\beta}_{\pmb{\gamma}}$ and $\pmb{\beta}_{\pmb{\gamma},-j}$ denotes all components of $\pmb{\beta}_{\pmb{\gamma}}$ except the $j$th component.  Furthermore, $\hat\sigma^2_{\pmb{\gamma}}$ and $\hat\sigma^2_{\pmb{\gamma^{\prime}}}$ are the MCELE of $\sigma^2_{\pmb{\gamma}}$ and $\sigma^2_{\pmb{\gamma^{\prime}}}$ 
corresponding to  $\pmb{\beta}_{\pmb{\gamma}}$ and $\pmb{\beta}_{\pmb\gamma^{\prime}}$ respectively.  In particular, using Theorem \ref{chapter2maxcond}, for a given  $\pmb{\beta}_{\pmb{\gamma}}$, the MCELE $\hat\sigma^2_{\pmb{\gamma}}$ is obtained by first maximising \eqref{eq:elms}
 under the constraints \eqref{eq:int}, \eqref{eq:model} and \eqref{eq:modelc} and then solving the equation in \eqref{eq:sigma} for $\sigma^2_{\pmb{\gamma}}$ after substituting $\pmb{\beta}_{\pmb{\gamma}}$ and the maximised weights appropriately.  Note that, by construction, $\sigma^2_{\pmb{\gamma^{\prime}}}$ is a function of both $\pmb{\beta}_{\pmb{\gamma}}$ and $u$. 



With these values of $(\pmb{\gamma^{\prime}},\pmb{\beta_{\gamma^{\prime}}}, \sigma^2_{\pmb{\gamma^{\prime}}})$, using \eqref{eq:postms}, we compute the BayesEL posterior $\Pi_{EL}\left(\pmb{\gamma},\pmb{\beta_{\gamma}},\sigma^{2}_{\pmb{\gamma}}\mid y,x\right)$ and accept this proposed state with probability 
\begin{equation}\label{alpha vary}
\alpha=\min\left\{1,\frac{\Pi_{EL}(\pmb{\gamma^{\prime}},\pmb{\beta_{\gamma^{\prime}}}, \sigma^2_{\pmb{\gamma^{\prime}}}\mid y,x)q_{\gamma}(\pmb{\gamma}\mid \pmb{\gamma}^{\prime})}{\Pi_{EL}(\pmb{\gamma},\pmb{\beta}_{\pmb{\gamma}}, \sigma^2_{\pmb{\gamma}}\mid y,x)q_{\gamma}(\pmb\gamma^{\prime}\mid \pmb\gamma)q_U(u)}\left|\frac{\partial (\pmb{\beta_{\gamma^{\prime}}}, \sigma^2_{\pmb{\gamma^{\prime}}})}{\partial (\pmb{\beta_{\gamma}}, u,\sigma^2_{\pmb{\gamma}})}\right|\right\},
\end{equation}

\begin{algorithm}[t]
\caption{Multi-step RJMCMC algorithm for BayesEL model selection.}
\label{algorithm2}

\begin{algorithmic}[1]


\REQUIRE $\left(\go,\pmb{\beta}_{\go},\sigma^{2}_{\go}\right)$.

\FOR {$t=1$ to $L$}

\STATE Compute $\hat{\pmb{\beta}}_{\gt}$;
\STATE Update within model: update parameters $\pmb{\beta}_{\gt}$, $\sigma^2_{\gt}$ in model $\gt$ and compute $\hat{\sigma}^2_{\gt}$ by two-step Metropolis Hastings method in Algorithm \ref{algorithm1};
\STATE Assign $\gtp=\gt$;
\STATE Propose an index $j$ randomly from $\{1,2,\ldots,s\}$ using a proposal $q_{\gamma}$; 
\IF{$\gt_j==0$}
\STATE $\gtp_{j}=1$;
\STATE Compute $\hat{\pmb{\beta}}_{\gtp}$;
\STATE Propose  $u$  from $q_U(\cdot)$\;
\STATE Compute
$(\pmb{\beta}_{\gtp,-j}, \pmb{\beta}_{\gtp,j})=\left(\pmb{\beta}_{\gt}+(\hat{\pmb{\beta}}_{\gtp,-j}-\hat{\pmb{\beta}}_{\gt}),u+\hat{\beta}_{\gtp,j}\right)$;
 


\STATE Compute $\hat{\sigma}^2_{\gtp}$ and $\sigma^2_{\gtp}=\sigma^2_{\gt}+(\hat\sigma^2_{\gtp}-\hat\sigma^2_{\gt})$;
\STATE Compute the empirical likelihood $\mc{L}(\gtp, \pmb{\beta}_{\gtp}, \sigma^2_{\gtp})$;

\STATE Accept the state $(\gtp, \pmb{\beta}_{\gtp}, \sigma^2_{\gtp})$ with probability 
{\small
\[
\alpha=\min\left\{1,\frac{\Pi_{EL}(\gtp,\pmb{\beta}_{\gtp}, \sigma^2_{\gtp}\mid y,x)q_{\gamma}(\gt|\gtp)}{\Pi_{EL}(\gt, \pmb{\beta}_{\gt}, \sigma^2_{\gt}\mid y,x)q_{\gamma}(\gtp\mid \gt)q_U(u)}\right\};
\]
}

\ELSE


\STATE $\gtp_j=0$;
\STATE Compute $\hat{\pmb{\beta}}_{\gtp}$, $\hat{\sigma}^2_{\gtp}$ as defined above;


\STATE $\pmb{\beta}_{\gtp}=\pmb{\beta}_{\gt,-j}+(\hat{\pmb{\beta}}_{\gtp}-\hat{\pmb{\beta}}_{\gt,-j})$;




\STATE Compute $\hat{\sigma}^2_{\gtp}$ and set $\sigma^2_{\gtp}=\sigma^2_{\gt}+(\hat{\sigma}^2_{\gtp}-\hat{\sigma}^2_{\gt})$;
\STATE Compute the empirical likelihood $\mc{L}(\gtp,\pmb{\beta}_{\gtp}, \sigma^2_{\gtp})$;
\STATE Accept the state $(\gtp, \pmb{\beta}_{\gtp}, \sigma^2_{\gtp})$ is accepted with probability 
{\small
\[
\alpha=\min\left\{1,\frac{\Pi_{EL}(\gtp,\pmb{\beta}_{\gtp}, \sigma^2_{\gtp}\mid y,x)q_{\gamma}(\gt|\gtp)q_{U}(\beta_{\gt,j}-\hat\beta_{\gt,j})}{\Pi_{EL}(\gt,\pmb{\beta}_{\gt}, \sigma^2_{\gt}\mid y,x)q_{\gamma}(\gtp\mid \gt)}\right\};
\]
}
\ENDIF
\ENDFOR
\RETURN $(\pmb{\gamma},\pmb{\beta}_{\pmb{\gamma}},\sigma^{2}_{\pmb{\gamma}})^{(1:L)}$.
\end{algorithmic}
\end{algorithm}

where $q_{\gamma}(\pmb{\gamma^{\prime}}\mid \pmb{\gamma})$ is the probability of proposing the jump from model $\pmb\gamma$ to model $\pmb{\gamma^{\prime}}$, and 
the last factor in \eqref{alpha vary} is the determinant of the Jacobian of the transformation from $(\pmb{\beta_{\gamma}}, u,\sigma^2_{\pmb{\gamma}})$ to $(\pmb{\beta_{\gamma^{\prime}}}, \sigma^2_{\pmb{\gamma^{\prime}}})$.

In order to jump from a $\pmb{\gamma}$ of dimension $k$ to a model $\pmb{\gamma}^{\prime}$ of dimension $k^{\prime}=k-1$, we note that the transformation $g_{\gamma\to\gamma^{\prime}}$ is injective, so its inverse can be used to map the parameters.  The new model and the parameters are then accepted with probability equal to the minimum of $1$ and $\alpha^{-1}$.

By construction it follows that $g_{\gamma\to\gamma^{\prime}}$ is a linear mapping and the determinant of Jacobian in \eqref{alpha vary} does not depend on the value of the parameters.

\begin{theorem}\label{chapter3prop1}
Consider the one-to-one mapping $g_{\gamma\to \gamma^{\prime}}$ as defined in equation \eqref{eq:g}.  Then the determinant of Jacobian in \eqref{alpha vary} is given by: 
\[
\bigg|\frac{\partial (\pmb{\beta}_{\pmb{\gamma}^{\prime}}, \sigma^2_{\pmb{\gamma}'})}{\partial(\pmb{\beta_\gamma}, u,\sigma^2_{\pmb{\gamma}})}\bigg|=1.
\]
\end{theorem}

The proposed RJMCMC algorithm is described in Algorithm \ref{algorithm2}.  Both stages of the algorithm are extensions of the two-stage Metropolis Hastings described in Section \ref{section fixed}.
 
In the first stage, where we update the parameters within a model $\gamma$, we first propose a new value of $\pmb{\beta}_{\pmb{\gamma}}$. A new value of $\sigma^2_{\pmb{\gamma}}$ is then proposed by directly applying Theorem \ref{chapter2maxcond} to obtain the MCELE $\hat{\sigma}^2_{\pmb{\gamma}}$ and generating a random value from a pre-specified proposal distribution depending on this MCELE.  
The new value of $\pmb{\beta}_{\pmb{\gamma}}$ can be proposed based on either its current value or the ordinary least squares estimator of the regression parameter for the model $\pmb{\gamma}$ ie.  $\hat{\pmb{\beta}}_{\pmb{\gamma}}$.  
In Algorithm \ref{algorithm2} (see Figure \ref{fig:rjmcmc}) we use the current value of $\pmb{\beta}_{\pmb{\gamma}}$ for this purpose.
Note that, one can view $\hat{\pmb{\beta}}_{\pmb{\gamma}}$ as a MCELE given the model $\pmb{\gamma}$ as well.  This MCELE however, is obtained by only imposing the score constraints in \eqref{eq:model} on the weights defining the empirical likelihood.      

The second stage of the algorithm, the injective mapping between the parameters of two models $\pmb{\gamma}$ and $\pmb{\gamma}^{\prime}$ depends on the above MCELEs of the respective models.  It should be noted that, for any index $j$ such that $\pmb{\gamma}_j=\pmb{\gamma}^{\prime}_j=1$, the relation, $\pmb{\beta}_{\pmb{\gamma},j}-\hat{\pmb{\beta}}_{\pmb{\gamma},j}=\pmb{\beta}_{\pmb{\gamma}^{\prime},j}-\hat{\pmb{\beta}}_{\pmb{\gamma}^{\prime},j}$ 
holds.  Furthermore, by construction, $\sigma^2_{\pmb{\gamma}}-\hat{\sigma}^2_{\pmb{\gamma}}=\sigma^2_{\pmb{\gamma}^{\prime}}-\hat{\sigma}^2_{\pmb{\gamma}^{\prime}}$.
Such constructions are intentional.  They ensure that the proposed parameters in the two models are equidistant from their MCELEs.  Therefore, 
the values of BayesEL posteriors under the two models would be close.  Thus even though  supports of 
$\Pi_{EL}(\pmb{\gamma},\pmb{\beta}_{\pmb{\gamma}}, \sigma^2_{\pmb{\gamma}}\mid y,x)$ and $\Pi_{EL}(\pmb{\gamma}^{\prime},\pmb{\beta}_{\pmb{\gamma}^{\prime}}, \sigma^2_{\pmb{\gamma}^{\prime}}\mid y,x)$ may be quite different, 
the proposed model would have a better chance of getting accepted.         

\section{Illustrative Applications}\label{sec:appl}
\subsection{Simultaneous Quantile Regression}\label{sec:multQuant}
Quantile regression, introduced by \cite{Koenker1978Bassett} tries to estimate the conditional quantile functions i.e. conditional distribution of the response variable expressed as a function of observed covariates. Suppose $Y$ is the response and $X$ is a vector of covariates, in quantile regression the $\tau^{th}$ conditional quantile of $Y$ given $X=x$, denoted by $\mathcal{Q}_\tau(Y|X)$ is modeled as 
\begin{equation}\label{eq:quantile}
\mathcal{Q}_\tau(Y|X)=X^T\theta(\tau)
\end{equation}

In this simulation study we consider a simple heteroskedastic regression model. Several conditional quantiles of the response is simultaneously estimated using a BayesEL framework.  Algorithm \ref{algorithm1} will be used to sample from the BayesEL posterior of the regression coefficients.

\begin{figure}[t]
  \begin{minipage}[c]{.35\columnwidth}
    \subfigure[Posterior means and standard deviations\label{tab:resQ}]{
      \centering
    \begin{tabular}{c||cc}
    parameter&mean&se\\\hline\hline
    $\theta_0\left(0.50\right)$&1.000&0.002\\
    $\theta_1\left(0.50\right)$&0.623&0.008\\
    $\theta_0\left(0.90\right)$&1.001&0.001\\
    $\theta_1\left(0.90\right)$&0.498&0.007\\
    $\theta_0\left(0.95\right)$&1.000&0.006\\
    $\theta_1\left(0.95\right)$&0.669&0.001\\\hline
    \end{tabular}
    }
    \subfigure[Trace plot of the $\theta_0\left(.95\right)$ \label{fig:trace}]{
      \resizebox{1.25\columnwidth}{!}{\includegraphics{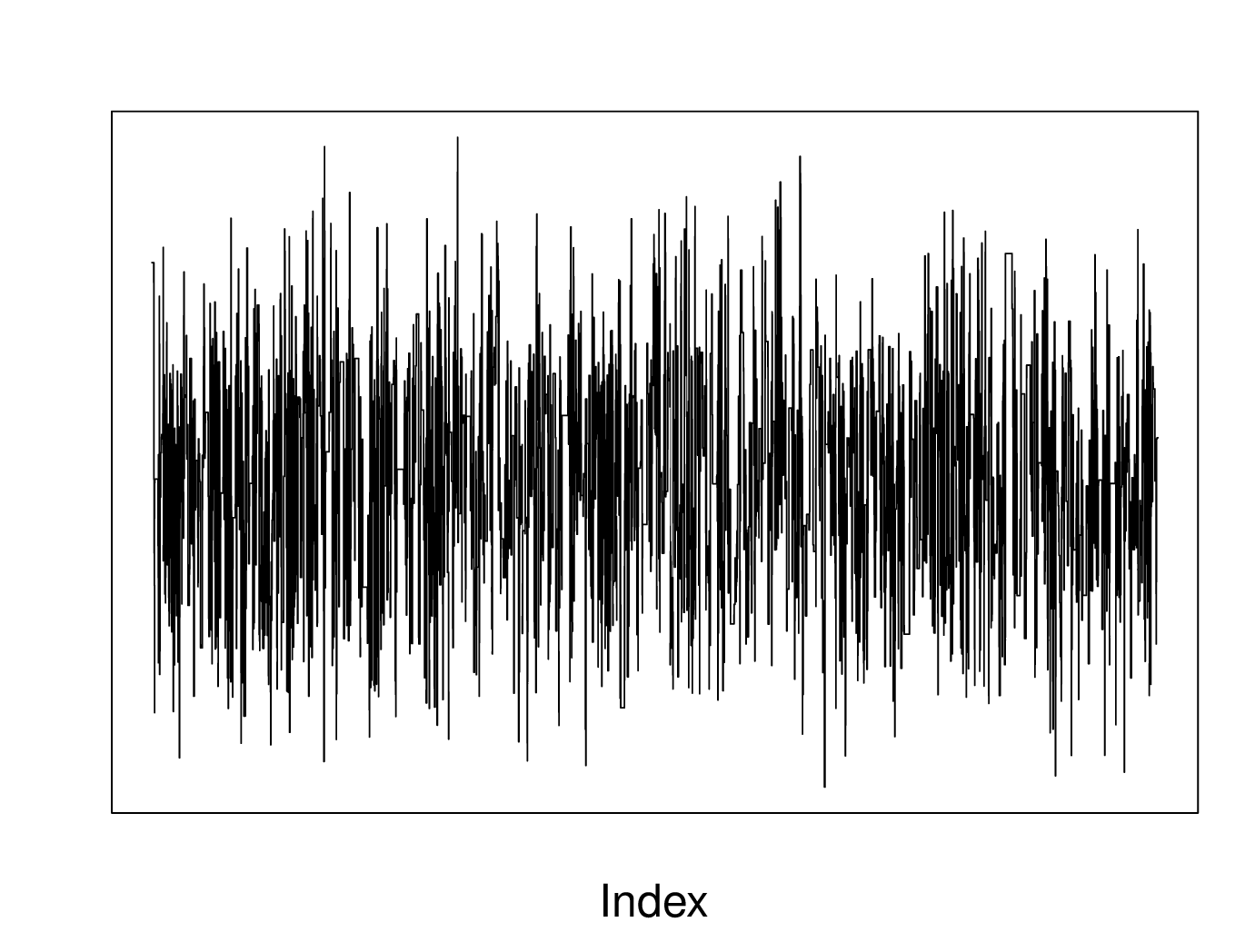}}
    }
  \end{minipage}\hfill
  \begin{minipage}[c]{.55\columnwidth}  
  \centering
  \subfigure[Fitted quantile regression lines. \label{fig:resQ}]{
    \resizebox{\columnwidth}{!}{\includegraphics{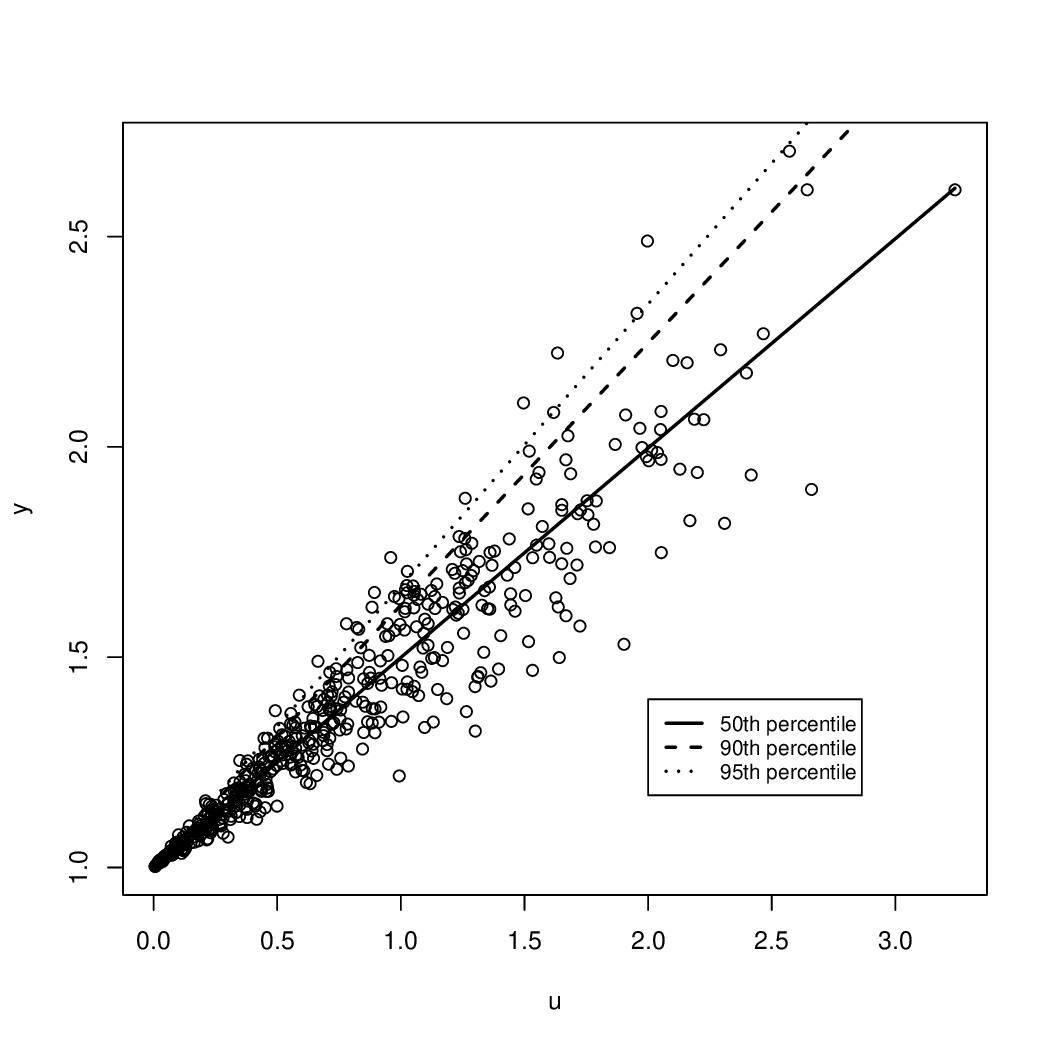}}
  }
  \end{minipage}
\caption{Some results from the BayesEL simultaneous quantile regression problem. \ref{tab:resQ} Posterior mean and standard deviations of the regression parameters, \ref{fig:trace} the trace plot of the intercept of the $.95^{th}$ quantile, and \ref{fig:resQ} the three fitted quantile regression lines on the data scatterplot.}
\label{fig:quantileReg}
\end{figure}

A sample $\left(u_i,y_i\right)$, $i=1$, $2$, $\ldots$, $500$, of size $n=500$, of i.i.d. observations was generated from the model:
\[
Y_i=1+\frac{1}{2}U_i+\epsilon_i,
\]

where $Y$ is the response variable, the covariate $U_i \sim \mid \mathcal{N}(0,1)\mid$ and the error $\epsilon_i\mid U_i=u_i \sim \mathcal{N}(0,0.01u_i^2)$, $i=1$, $2$, $\ldots$, $n$.  The chosen heteroskedastic model is particulary suitable for this illustration.  This is because the the regression lines for different values of $\tau$ would not be parallel to each other (See Figure \ref{fig:quantileReg}).  In this simulation study, we estimate $\mathcal{Q}_\tau(Y|X)$ simultaneously for $\tau=\left(\tau_1,\tau_2,\tau_3\right)^T=\left(.50,.90,.95\right)^T$. 

Quantile regression in Bayesian paradigm has been studied by several authors before (eg. \citet{dasGhoshal2017}, \citet{dasGhoshal2018}).  Empirical likelihood provides an alternative way to specify a likelihood for the regression coefficients. BayesEL-based quantile regression was studied by \citet{yang2012bayesian} and in small area estimation by \citet{chenLiu2019}.   
  

  The empirical likelihood-based formulation of the quantile regression problem directly follows from its construction.  Let for $i=1$, $2$, $\ldots$, $n$, $X_i=[1,U_i]$.  For $j=1,2,3$, define $\theta(\tau_j)=\left[\theta_0(\tau_j),\theta_1(\tau_j)\right]^T$, $\psi_{\tau_j}(u) =1_{\{u<0\}}-\tau_j$.  Furthermore, for each $i$ and $j$ define:
  \[
  m_{j}\left(X_i,Y_i,\theta(\tau_j)\right)=\psi_{\tau_{j}}(Y_i-\theta_0(\tau_j)-U_i\theta_1(\tau_j))[1,U_i]=\psi_{\tau_{j}}(Y_i-X_i\theta(\tau_j))X_i,
  \]
  and
  \[
  m_i\left(\theta(\tau_1),\theta(\tau_2),\theta(\tau_3)\right)=\left[m_{1}\left(X_i,Y_i,\theta(\tau_1)\right),m_{2}\left(X_i,Y_i,\theta(\tau_2)\right),m_{3}\left(X_i,Y_i,\theta(\tau_3)\right)\right]^T.
  \]
  Note that, for each $i=1$, $2$, $\ldots$, $n$, $m_i\left(\theta(\tau_1),\theta(\tau_2),\theta(\tau_3)\right)$ is a six dimensional vector of functions.  For quantile regression the empirical likehood for $\left(\theta(\tau_1),\theta(\tau_2),\theta(\tau_3)\right)$ is obtained by solving \eqref{el} under the constraint:
  \[
  \mathcal{W}\left(\theta(\tau_1),\theta(\tau_2),\theta(\tau_3)\right)=\left\{w~:~\sum^n_{i=1}w_im_i\left(\theta(\tau_1),\theta(\tau_2),\theta(\tau_3)\right)=0\right\}\cap\Delta_{n-1}
  \]
  
From this the BayesEL formulation in \eqref{posterior} follows trivially.  We assume independent noninformative $\mathcal{N}(0,100^2)$ priors on each of the six regression parameters.
    

Sampling from the resulting BayeEl posterior is complicated for two reasons. First, the six regression coefficient are highly dependent in their joint posterior, which makes proposing an acceptable value difficult.  Second, the empirical likelihood and as a result the BayesEL posterior is a piece-wise constant function with respect to the coefficients.  A platikurtic distribution which locally represents the flatness of the likelihood would be ideal for a proposal distribution.

The first problem is addressed by our proposed method.  By denoting $x_i=[1,u_i]$, for $i=1$, $2$, $\ldots$, $n$, we device a two-step procedure as described in Algorithm \ref{algorithm1} with:

\begin{equation*}
  \mathcal{W}_{G}(\theta(\tau_1))=\Bigg\{\omega:\sum_{i=1}^{n}\omega_i\left[m_1\left(x_i,y_i,\theta(\tau_1)\right)\right]^T=0\Bigg\},
\end{equation*}

and

\begin{equation*}
  \mathcal{W}_{H}(\theta(\tau_2),\theta(\tau_3))=\Bigg\{\omega:\sum_{i=1}^{n}\omega_i\left[m_2\left(x_i,y_i,\theta(\tau_2)\right),m_3\left(x_i,y_i,\theta(\tau_3)\right)\right]^T=0\Bigg\}.
\end{equation*}


To address the issue of piece-wise flatness of the posterior, a platikurtic distribution, e.g. the exponential power distribution is chosen as proposal distribution.  The denisty is given by \citep{johnson1987simulation}:
\begin{equation}\label{eq:quantQ}
q(x)=\frac{1}{2\alpha^{(1/\alpha)} \Gamma(1+1/\alpha)\sigma_q} e^{-\frac{|x-\mu_q|^{\alpha} }{\alpha\sigma_q^{\alpha}}},
\end{equation}
where $\mu_q$ is location parameter, $\sigma_q$ is the scale parameter and $\alpha$ the shape parameter.  Note that, the proposal density is supported over the whole real line, which ensures ergodicity.  The platikurticity increases with the shape parameter $\alpha$.   
  
In this example we take $\tau_1=.50$, $\tau_2=.90$, $\tau_3=.95$.  A sample of size $250,000$ were drawn from the posterior of $\theta(\tau_1)$, $\theta(\tau_2)$, and $\theta(\tau_3)$. An acceptance rate of $19.33\%$ was achieved with appropriate choices of shape and scale parameters of $q(x)$ in \eqref{eq:quantQ}.  

Some results, based on the last $125,000$ samples are presented in Figure \ref{fig:quantileReg} above.  For the fitted quantile regression lines, posterior means were used for estimating the regression parameters. The traceplots on the left panels are for the intercept and slope for the $.95^{th}$ percentile (ie. $\theta\left(\tau_3\right)$).  More results can be found in Section \ref{sec:multQuantSupp} of the Supplement.

\subsection{Gene Expression Data}\label{sec:gene}

We now turn to BayesEL model selection and present an example in Graphical Markov model selection.  Expression of $40$ genes in the MVA and MEP pathways of \textit{Arabidopsis thaliana} were collected from $118$ microarray assays.  Among them 
the pathway structure between $13$ genes on the MEP pathway was studied by \citet{wille2004sparse}.  They used a modified Gaussian graphical modelling approach to select the interaction structure.  
\citet{drton2007multiple} employed a multiple testing based graphical model selection to select a directed acyclic graph among the genes.  However, they also assumed that the expressions are normally distributed.  
In this example we examine the same data-set, but use our BayesEL approach for model selection and employ the proposed RJMCMC algorithm to sample from the resulting posterior.


Any viable strategy of directed acyclic graph selection requires one to first specify an order among the variables.  The selection procedure primarily chooses the parents (ie. the \emph{parenthood}) of 
a particular node from the nodes which precedes it in the ordering.  Justification for this procedure can be derived from the theory of Graphical Markov models for directed acyclic graphs \citep{lauritzen_book_1996}.  
Furthermore, it can be shown that the joint data likelihood factors according to the graph, which means each that the parents of each node can be selected independent of the others. 


In this example, we assume the same order in the genes as in \citet{drton2007multiple}.  Like them, we also assume that the first three nodes don't have any parents.  So we select parenthoods of rest of the ten nodes.    
 
For $k\in\{4,5,\ldots,13\}$ suppose $g_k$ denotes the $k^{th}$ gene. Given $g_k$, the model $\pmb{\gamma}_{(k)}=(\gamma_1,\ldots,\gamma_{k-1})$, is fit to the gene.  Here $\gamma_i$ is 1 when the gene $g_i$ is in the model and 0 otherwise. Let $g_{\pmb{\gamma}_{(k)}}$ 
be the matrix of covariates with columns $g_{i}$ such that $\pmb{\gamma}_{(k)}=1$, $\beta_{\pmb{\gamma}_{(k)}}$ be the corresponding vector of regression coefficients. As in \eqref{linear}, the model $\pmb{\gamma}_{(k)}$ for $g_k$ in terms of its parents is then given by:

\begin{equation}
E\left[g_k\mid g_{\pmb{\gamma}_{(k)}}\right] = g_{\pmb{\gamma}_{(k)}}\beta_{\pmb{\gamma}_{(k)}}\qquad\text{and}\qquad Var\left[g_k\mid g_{\pmb{\gamma}_{(k)}}\right] = \sigma^2_{\pmb{\gamma}_{(k)}}.
\end{equation}

We assume a double exponential $(0, \lambda)$ and an inverse gamma $(0.1,0.1)$ prior for $\beta_{\pmb{\gamma}_{(k)}}$ and $\sigma_{\pmb{\gamma}_{(k)}}^2$ respectively.  The hyper-parameter $\lambda$ is assumed to follow an inverse gamma $(5,5)$ prior.  
Our choice of double exponential priors for the regression parameters mimics the $L_1$ penalisation in LASSO \citep{tibshirani_1994}.  The parameter $\lambda$ controls the stringency of this penalty. In our setup, the amount of shrinkage differs from one parenthood to another. 
Each $\pmb{\gamma}_i$ is assumed to be a Bernoulli random variable, where the success probability follows a Beta $(2,7)$ distribution.  Such priors would prefer sparser models.

The constraints imposed on \eqref{eq:elms} to compute the empirical likelihood of the parenthood of the $k$th gene can be easily specified from equations \eqref{eq:int}, \eqref{eq:model}, \eqref{eq:modelc}, \eqref{eq:sigma} and \eqref{eq:wms}.  
The observation for the $k$th gene i.e. $g_k$ is used as response and each of the genes $g_i$, $i=1$, $2$, $\ldots$, $k-1$ a possible covariate.  The standardised response and the covariates are used.  The set of constraints on the weights for the parenthood of the $k$th gene is given by:     
\[
\mathcal{W}\left(\pmb{\gamma}_{(k)},\pmb{\beta}_{\pmb{\gamma}_{(k)}},\sigma^2_{\pmb{\gamma}_{(k)}}\right)=\mathcal{W}_{0}\left(\pmb{\beta}_{\pmb{\gamma}_{(k)}}\right)\bigcap\mathcal{W}_{\gamma_{(k)}}\left(\pmb{\beta}_{\pmb{\gamma}_{(k)}}\right)\bigcap\mathcal{W}_{\gamma^c_{(k)}}\left(\pmb{\beta}_{\pmb{\gamma}_{(k)}}\right)\bigcap\mathcal{W}_{\sigma}\left(\pmb{\beta}_{\pmb{\gamma}_{(k)}},\sigma^2_{\pmb{\gamma}_{(k)}}\right).
\]



\begin{table}[t]
  \begin{center}
    \begin{tabular}{p{.09\columnwidth}|>{\raggedright\arraybackslash}p{.275\columnwidth}p{.05\columnwidth}||p{.09\columnwidth}|>{\raggedright\arraybackslash}p{.275\columnwidth}p{.05\columnwidth}}\hline
  node&parents&prob&node&parents&prob\\\hline\hline
  DXR&DXPS2, DXPS3&0.904&MCT&DXR&0.538\\
  &DXPS2&0.044&&DXPS2, DXR&0.391\\\hline
  CMK&DXPS3, MCT&0.226&MECPS&DXPS1, DXPS3, DXR&0.267\\
  &MCT&0.185&&DXPS1, DXPS3, DXR, CMK&0.264\\\hline
  HDS&DXPS1, MECPS&0.220&HDR&DXPS2, MCT&0.187\\
  &DXPS1, DXPS2, DXR, MECPS&0.137&&DXPS2, CMK&0.154\\\hline
  IPPI1&HDS&0.209&GPPS&DXPS1, DXPS2, HDR&0.285\\
  &MECPS&0.132&&DXPS1, DXPS2, DXR, HDR&0.104\\\hline
  PPDS1&GPPS, HDR&0.386&PPDS2&DXPS2, DXR, CMK, IPPI1, PPDS1&0.239\\
  &DXPS1, GPPS, HDR&0.235&&DXPS2, DXR, IPPI1, PPDS1&0.223\\\hline\hline
  \end{tabular}

    \end{center}
  \caption{The node-wise models with highest Monte Carlo probabilities obtained from the proposed RJMCMC algorithm.}
  \label{tab:mods}
\end{table}

Now given $\pmb{\gamma}_{(k)}$, $\pmb{\beta}_{\pmb{\gamma}_{(k)}}$, $\sigma^2_{\pmb{\gamma}_{(k)}}$ the empirical likelihood is given by

\[
\mc{L}(\pmb{\gamma}_{(k)}, \pmb{\beta}_{\pmb{\gamma}_{(k)}}, \sigma^2_{\pmb{\gamma}_{(k)}}) = \max_{\omega\in\mathcal{W}\left(\pmb{\gamma}_{(k)}, \pmb{\beta}_{\pmb{\gamma}_{(k)}}, \sigma^2_{\pmb{\gamma}_{(k)}}\right)}\prod_{i=1}^n \omega_i.
\]
The likelihood is zero if the maximisation problem is infeasible.
From the likelihood the BayesEL posterior for $\pmb{\gamma}_{(k)}$, $\pmb{\beta}_{\pmb{\gamma}_{(k)}}$, $\sigma^2_{\pmb{\gamma}_{(k)}}$ can be computed.  
 

Given model ${\pmb{\gamma}_{(k)}}$, the proposal distribution for $\beta_{\pmb{\gamma}_{(k)}}$ is Gaussian distribution with its current value as mean and $0.03$ as standard deviation. For proposing $\sigma^2_{\pmb{\gamma}_{(k)}}$, truncated normal with its MCELE as mean and $1$ as standard deviation is used. 
In the case that a covariate is added, the value of $\beta_{\pmb{\gamma}_{(k)}}$ is proposed from $\mathcal{N}(0,0.0025)$.  With these proposals, samples of size $500,000$ are drawn from the derived BayesEL posterior and the first $400,000$ is discarded as burn-in.




\begin{figure}[t]
\centering
\subfigure[\label{fig:md}]{
\begin{tikzpicture}

\tikzset{vertex/.style = {shape=ellipse,draw,scale=0.7}}
\tikzset{edge/.style = {->,> = latex'}}
\node[vertex] (MECPS) at  (0,0) {MECPS};
\node[vertex] (CMK) at  (0,1) {CMK};
\node[vertex] (MCT) at  (0,2) {MCT};
\node[vertex] (DXR) at  (0,3) {DXR};
\node[vertex] (DXPS2) at (0,4) {DXPS2};
\node[vertex] (DXPS1) at (-2,4) {DXPS1};
\node[vertex] (DXPS3) at (2,4) {DXPS3};
\node[vertex] (HDS) at (0,-1) {HDS};
\node[vertex] (HDR) at (0,-2) {HDR};
\node[vertex] (IPPI1) at (0,-3) {IPPI1};
\node[vertex] (PPDS1) at (0,-4) {PPDS1};
\node[vertex] (GPPS) at (-2,-4) {GPPS};
\node[vertex] (PPDS2) at (2,-4) {PPDS2};
\draw[edge] (DXPS2) to (DXR);
\draw[edge] (DXPS3) to (DXR);
\draw[edge] (DXR) to (MCT);
\draw[edge] (MCT) to (CMK);
\draw[edge] (MECPS) to (HDS);
\draw[edge] (HDR) to (GPPS);
\draw[edge] (GPPS) to (PPDS1);
\draw[edge] (PPDS1) to (PPDS2);
\draw[edge] (IPPI1)  to (PPDS2);

\draw[edge] (DXR)  to[bend left=40] (MECPS);
\draw[edge] (DXPS3)  to[bend left] (MECPS);
\draw[edge] (DXPS1)  to[bend right] (HDS);
\draw[edge] (DXPS2)  to[bend right=50] (HDS);
\draw[edge] (MCT)  to[bend right=50] (HDS);
\draw[edge] (DXR)  to[bend left=45] (HDS);
\draw[edge] (DXPS2)  to[bend right=50] (HDR);
\draw[edge] (HDS)  to[bend left=50] (IPPI1);
\draw[edge] (HDR)  to[bend right=50] (PPDS1);
\draw[edge] (DXR)  to[bend left] (PPDS2);

%
%
%
\end{tikzpicture}}
\subfigure[\label{fig:yt}]{
\begin{tikzpicture}

\tikzset{vertex/.style = {shape=ellipse,draw,scale=0.7}}
\tikzset{edge/.style = {->,> = latex'}}
\node[vertex] (MECPS) at  (0,0) {MECPS};
\node[vertex] (CMK) at  (0,1) {CMK};
\node[vertex] (MCT) at  (0,2) {MCT};
\node[vertex] (DXR) at  (0,3) {DXR};
\node[vertex] (DXPS2) at (0,4) {DXPS2};
\node[vertex] (DXPS1) at (-2,4) {DXPS1};
\node[vertex] (DXPS3) at (2,4) {DXPS3};
\node[vertex] (HDS) at (0,-1) {HDS};
\node[vertex] (HDR) at (0,-2) {HDR};
\node[vertex] (IPPI1) at (0,-3) {IPPI1};
\node[vertex] (PPDS1) at (0,-4) {PPDS1};
\node[vertex] (GPPS) at (-2,-4) {GPPS};
\node[vertex] (PPDS2) at (2,-4) {PPDS2};
\draw[edge] (DXPS2) to (DXR);
\draw[edge] (DXPS3) to (DXR);
\draw[edge] (DXR) to (MCT);
\draw[edge] (MCT) to (CMK);
\draw[edge] (DXPS3)  to[bend left] (CMK);
\draw[edge] (DXR)  to[bend left=40] (MECPS);
\draw[edge] (DXPS3)  to[bend left] (MECPS);
\draw[edge] (DXPS1)  to[bend right] (MECPS);
\draw[edge] (MECPS) to (HDS);
\draw[edge] (DXPS2)  to[bend right=40] (HDS);
\draw[edge] (DXPS2)  to[bend right=40] (HDR);
\draw[edge] (MCT)  to[bend left=50] (HDR);
\draw[edge] (HDS)  to[bend left=50] (IPPI1);
\draw[edge] (HDR) to (GPPS);
\draw[edge] (DXPS1)  to[bend right=20] (GPPS);
\draw[edge] (DXPS2)  to[bend right=25] (GPPS);
\draw[edge] (GPPS) to (PPDS1);
\draw[edge] (HDR)  to[bend right=50] (PPDS1);
\draw[edge] (PPDS1) to (PPDS2);
\draw[edge] (IPPI1)  to (PPDS2);
\draw[edge] (DXR)  to[bend left=30] (PPDS2);
\draw[edge] (DXPS2)  to[bend left=40] (PPDS2);
\draw[edge] (CMK)  to[bend left] (PPDS2);

%
%
%
\end{tikzpicture}}
\caption{Directed acyclic graphic selected by (a) the step-down Sidak procedure when family-wise error rate is controlled at $0.1$ and (b) by two-step reversible jump MCMC with Bayesian empirical likelihood.}
\label{Graphic_model_191114}
\end{figure}

In Table \ref{tab:mods} we present the top two set of parents for each node according to the Monte Carlo probability as obtained from the proposed RJMCMC procedure.  The set of parents of each node is taken to be the model with highest Monte Carlo probability in the above sample.  The selected directed acyclic graph is shown in Figure \ref{Graphic_model_191114}.  We compare our results (Figure \ref{fig:yt})with the graph selected by \citet{drton2007multiple} (Figure \ref{fig:md}), using a step-down Sidak procedure where the family-wise error rate (FWER) is controlled at $0.1$.  
The step-down method chooses a graph with $19$ edges, which is sparser than the graph we obtained with $23$ edges. The structure of metabolic network obtained from the two approaches are similar.   It can be seen that, the parents of DXR, MCT, IPPI1 and PPDS1, are same in for both methods. 
 All but one arrows (DXPS1 $\rightarrow$ HDS) selected by the step-down method are selected by our proposed method as well.  

The proposed RJMCMC sampler appears to move between the models well.  The acceptances rates of the cross model moves were quite high for all nodes.  In fact, for all but one node, these rates were higher than 10\%.  For some it was even higher than 20\%. See Section \ref{sec:suppGene} in the supplement for more details.

\section{Discussion}
In this article we present a novel method of sampling from a BayesEL posterior with a possible non-convex support.  In a BayesEL procedure, instead of specifying a parametric likelihood of the data, one uses a likelihood obtained from a constrained estimate of the distribution function.  The constraints depend on the model and its parameters.
A non-convex support often makes sampling from a BayesEL posterior hard, which has prevented its wider use in statistical analysis.

The proposed method can be described as a two-step Metropolis-Hastings algorithm, where new values of an appropriate subset of parameters are proposed first. Next, using these values and the estimating equations we compute a maximum conditional empirical likelihood estimator (MCELE) of the rest of the parameters.  
New values of these parameters are then proposed close to their maximum conditional empirical likelihood estimates.  The proposed method does not require any smoothness of the estimating equations and can be used for non smooth and even discontinuous estimating functions.

We show that the proposed method can easily be extended to an appropriate reversible jump Markov chain Monte Carlo (RJMCMC) procedure, which would allow the efficient implementation of Bayesian model selection using BayesEL procedure.  
As far as we know, this is the first implementation of RJMCMC procedure on BayesEL posteriors.  Without the proposed two-step Metropolis-Hastings method, it would be almost impossible to use 
empirical likelihood in Bayesian model selection problems.

Assumptions made in the article can potentially be relaxed.  First, the assumption that equation \eqref{eqn of theta_2} has a unique solution is not required in many situations.  In the presence of multiple solutions, our method would work as described if one of the solutions could be chosen deterministically.  
This would apply to the proposed RJMCMC when the covariates are not independent. The ordinary least squares estimator can be deterministically specified by a specified choice of generalised inverse. 
If all the solutions can be computed either analytically or numerically.  We can even proceed by randomly choosing one of the solutions as our MCELE.  However in this case, the transition probability would depend on all solutions.  Most real-life problems should satisfy one of the above conditions.


The specific structure of the estimating equations described in Section \ref{section-max} is not strictly necessary.  The equation $h$ may depend only on one set of parameters (i.e. $\theta_2$) and be free of the other set (i.e. $\theta_1$) (see Section \ref{sec:multQuant}).  In such cases, instead of proposing the value $\theta_2$ based on the proposed value of $\theta_1$ one can execute the opposite, i.e.
propose a value of $\theta_1$ based on the proposed value of $\theta_2$.  
Even if $h$ depends both on $\theta_1$ and $\theta_2$, from a value of $\theta_2$, one can obtain a MCELE of $\theta_1$ (i.e. $\hat{\theta}_1\left(\theta_2\right)$), by direct maximisation of \eqref{max cond theta_2 2}. A value of $\theta_1$ can be subsequently proposed from $\hat{\theta}_1\left(\theta_2\right)$.
This MCELE however may not be unique. So some caution need to be exercised.  




\appendix 
\section{Proofs}
\noindent{\it Proof of Theorem \ref{chapter2maxcond}.}\label{appendix1}
Since $\hat\nu(a)$ maximises over $\mathcal{W}_G(a)$ and $\hat\nu(a)\in\mathcal{W}(a,\tilde{\theta}_2)$, by the uniqueness of the solution, we get $\hat\omega(a,\tilde\theta_2)=\hat\nu(a)$ and $L\left(a,\tilde{\theta}_2\right)=\prod^n_{i=1}\hat\nu_i(a)$.  By definition of $\Tet\left(\theta_1\right)$ it follows that,
\[
\prod^n_{i=1}\hat{\nu}_i(a)=L\left(a,\tilde{\theta}_2\right)\le L\left(a,\Tet\left(\theta_1\right)\right)=\prod^n_{i=1}\hat\omega_i\left(a,\Tet\left(\theta_1\right)\right)\quad\text{(say)}.
\] 

However, $\hat\omega\left(a,\Tet\left(\theta_1\right)\right)\in\mathcal{W}_G(a)$. That is
\[
\prod_{i=1}^n\hat{\omega}_{i}\left(a,\Tet\left(\theta_1\right)\right)=L\left(a,\Tet\left(\theta_1\right)\right)\le\prod^n_{i=1}\hat\nu_i(a)=L\left(a,\tilde{\theta}_2\right).
\]

The the statement holds.\hfill$\square$
\smallskip

\noindent{\it Proof of Theorem \ref{chapter3thm1}.}\label{appendix2}
From $\theta_1^{(t+1)}\ci \left(\theta_1^{(1:t-1)}, \theta_2^{(1:t)}\right)\big | \theta_1^{(t)}$, using the properties of the conditional independence \citep{dawid1979conditional},  it follows that
\begin{equation}\label{second}
\theta_1^{(t+1)}\ci \left(\theta_1^{(1:t-1)}, \theta_2^{(1:t-1)}\right)\big | \left(\theta_1^{(t)},\theta_2^{(t)}\right).
\end{equation}
From $\theta_2^{(t+1)}\ci \left(\theta_1^{(1:t)},\theta_1^{(1:t)}\right)\big | \theta_1^{(t+1)}$, similarly, it follows that
\begin{equation}\label{first}
\theta_2^{(t+1)}\ci \left(\theta_1^{(1:t-1)},\theta_1^{(1:t-1)}\right)\big|\left(\theta_1^{(t+1)},\theta_1^{(t)},\theta_2^{(t)}\right).
\end{equation}
From \eqref{first} and \eqref{second}, using the conditional independence again, we get
\[
\left(\theta_1^{(t+1)},\theta_2^{(t+1)}\right)\ci \left(\theta_1^{(1:t-1)},\theta_2^{(1:t-1)}\right)\mid\left(\theta_1^{(t)},\theta_2^{(t)}\right).
\]
Thus $\left(\theta_1^{(1:L)},\theta_2^{(1:L)}\right)$ is a Markov chain. 

The transition probability is
\[
P\left(\theta_1^{(t+1)},\theta_2^{(t+1)}\mid\theta_1^{(t)},\theta_2^{(t)}\right)=P\left(\theta_2^{(t+1)}\mid\theta_1^{(t+1)},\theta_1^{(t)},\theta_2^{(t)}\right)P\left(\theta_1^{(t+1)}\mid\theta_1^{(t)},\theta_2^{(t)}\right).
\]

By construction, $\theta_1^{(t+1)} \ci \theta_2^{(t)}\mid\theta_1^{(t)}$. Thus
$P\left(\theta_1^{(t+1)}\mid\theta_1^{(t)},\theta_2^{(t)}\right)=P\left(\theta_1^{(t+1)}\mid\theta_1^{(t)}\right)$.
Furthermore

\begin{eqnarray*}
&& P\left(\theta_2^{(t+1)}\mid\theta_1^{(t+1)},\theta_1^{(t)},\theta_2^{(t)}\right)=P\left(\theta_2^{(t+1)}\mid\theta_1^{(t+1)}\right)=\int P\left(\theta_2^{(t+1)},\tilde{\theta}_2\mid\theta_1^{(t+1)}\right)d\tilde{\theta}_2\\
&=&\int P\left(\theta_2^{(t+1)}\mid\tilde{\theta}_2,\theta_1^{(t+1)}\right)P\left(\tilde{\theta}_2\mid\theta_1^{(t+1)}\right)d\tilde{\theta}_2=P\left(\theta_2^{(t+1)}\mid\theta_1^{(t+1)},\Tet\left(\theta_1^{(t+1)}\right)\right).
\end{eqnarray*}

The last equality holds since $P\left(\tilde{\theta}_2\mid \theta_1^{(t+1)}\right)=1$  if $\tilde{\theta}_2=\Tet\left(\theta_1^{(t+1)}\right)$ and $0$ otherwise.  Therefore
\[
P\left(\theta_1^{(t+1)},\theta_2^{(t+1)}\mid\theta_1^{(t)},\theta_2^{(t)}\right)=P\left(\theta_1^{(t+1)}\mid\theta_1^{(t)}\right)P\left(\theta_2^{(t+1)}\mid \theta_1^{(t+1)},\Tet\left(\theta_1^{(t+1)}\right) \right).
\]\hfill$\square$
\smallskip

\noindent{Proof of Theorem \ref{chapter3prop1}.}\label{appendix3}
By construction $\pmb{\beta}_{\pmb{\gamma}^{\prime}}$ does not depend on $\sigma^2_{\pmb{\gamma}^{\prime}}$.  However, the later depends on $\pmb{\beta}_{\pmb{\gamma}}$ and $u$ through $\hat{\sigma}^2_{\pmb{\gamma}^{\prime}}$. Thus,
\[
\left|\frac{\partial (\pmb{\beta}_{\pmb{\gamma}^{\prime}}, \sigma^2_{\pmb{\gamma}'})}{\partial(\pmb{\beta_\gamma}, u,\sigma^2_{\pmb{\gamma}})}\right|=\left|\frac{\partial g_{\gamma\to \gamma^{\prime}}(\pmb{\beta_\gamma}, u,\sigma^2_{\pmb{\gamma}})}{\partial(\pmb{\beta_\gamma}, u,\sigma^2_{\pmb{\gamma}})}\right|=\left|\begin{array}{ccc}
1 & 0 & ~~~~~0~~~~~ \\
0 & 1& ~~~~~0 ~~~~~\\
\frac{\partial \hat\sigma^2_{\pmb{\gamma^{\prime}}}}{\partial\pmb{\beta_{\gamma}}} & \frac{\partial \hat\sigma^2_{\pmb{\gamma^{\prime}}}}{\partial u} & ~~~~~1~~~~~
\end{array}
\right|=1.
\]\hfill$\square$

\section{Simultaneous Quantile Regression}\label{sec:multQuantSupp}
In this section, we present additional results from the simultaneous quantile regression described in Section \ref{sec:multQuant} in the main text. 

\begin{figure}[h]
  \resizebox{\columnwidth}{!}{\includegraphics{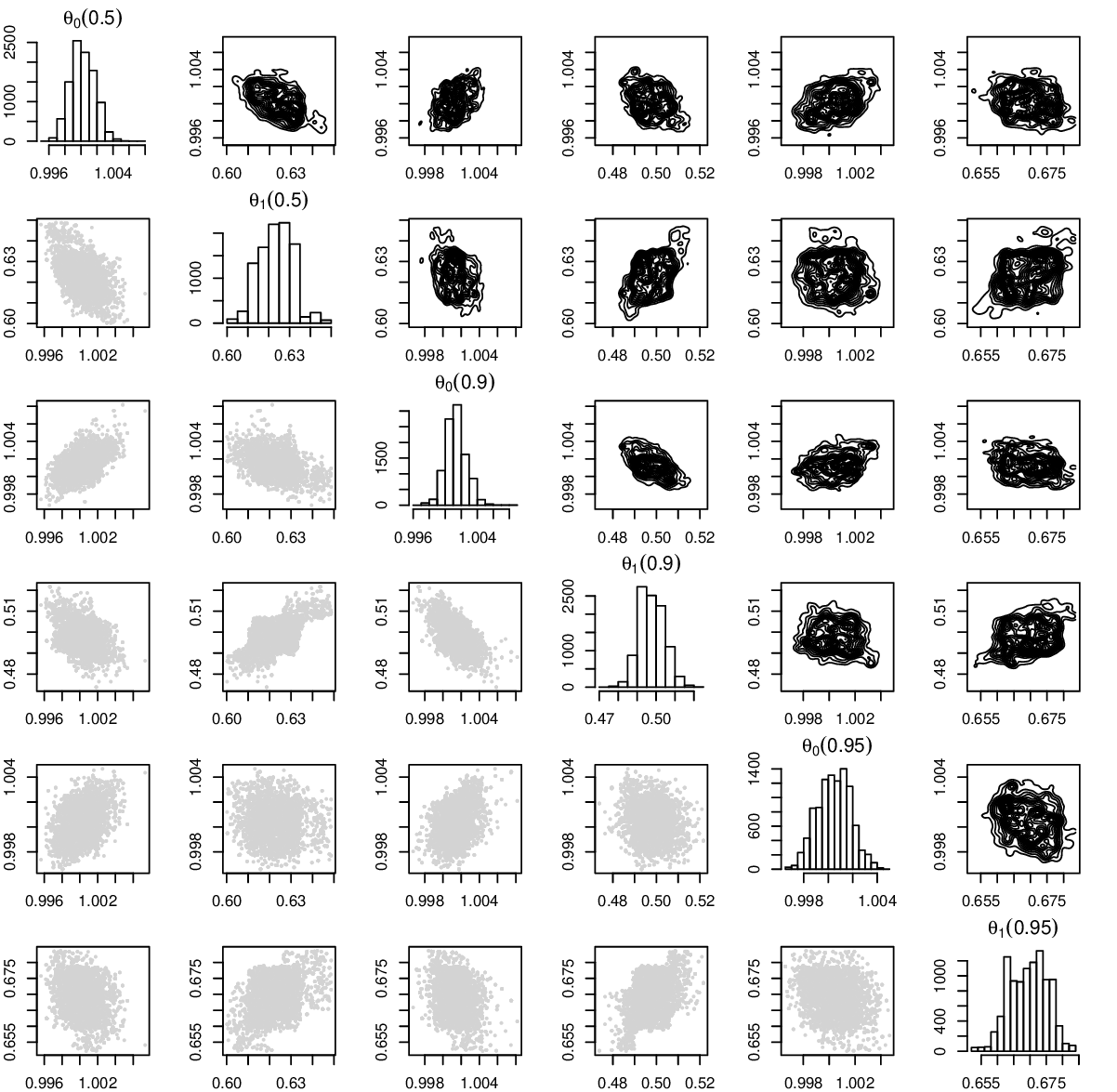}}
  \caption{The sample scatter and contours of the pairwise BayesEl posterior density of the regression parameters for the quantile regression for $\tau_1=0.50$, $\tau_2=.90$, and $\tau_3=0.95$. The histogram of the diagonal represents the BayesEl posterior marginal densities of the parameters.}
  \label{fig:cNs}
\end{figure}

In Figure \ref{fig:cNs} we present the marginal and pairwise pairwise distributions of the six regression parameters.  The marginal density of the parameters are presented in the diagonal of the figure.  All posteriors are unimodal and might be normally distributed. The pairwise scatter and contours of the two-dimensional density plots are presented in the diagonal. The pairwise dependence of the regression parameters in the posterior is evident.     

\begin{figure}[t]
  \resizebox{\columnwidth}{!}{\includegraphics{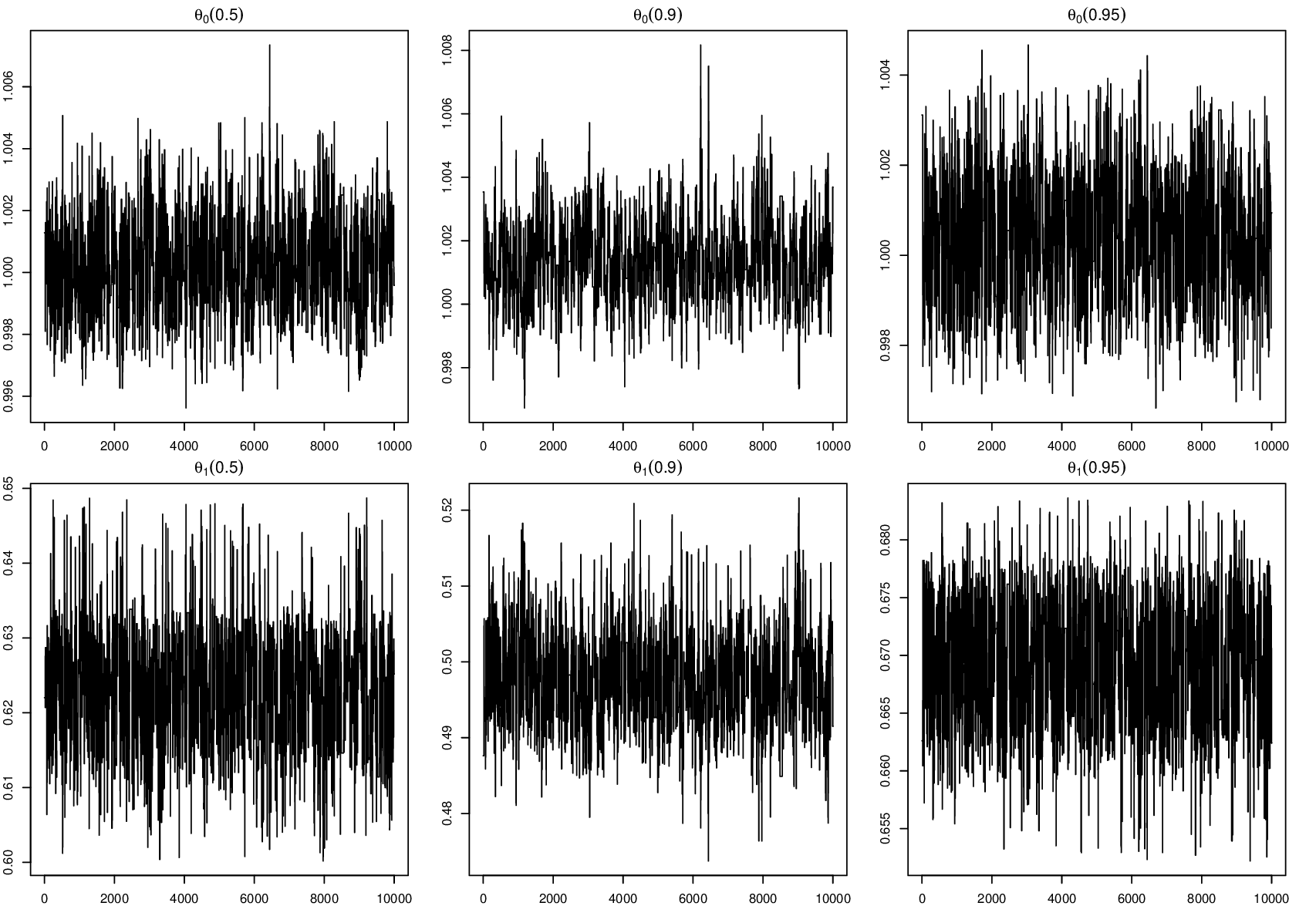}}
  \caption{Trace-plots of the quantile regression parameters from the proposed MCMC sampling.}
  \label{fig:trcplt}
\end{figure}

In Figure \ref{fig:trcplt} we present the marginal trace-plots from the proposed posterior MCMC sampling procedure of the six regression parameters.  The figures show that the chains mix well, and the proposed procedure works. 

\section{Gene Expression Data}\label{sec:suppGene}
In this section we present results from some diagnostic tests for the directed acyclic graph selection for the gene expression data in Section \ref{sec:gene}.

\begin{table}[h]
  \begin{center}
  \begin{tabular}{p{2cm}||l|l|l|l|l}\hline
    node&DXR&MCT&CMK&MECPS&HDS\\
    acceptance rate&0.066&0.213&0.273&0.150&0.114\\\hline
    node&HDR&IPPI1&GPPS&PPDS1&PPDS2\\
    acceptance rate&0.202&0.185&0.173&0.144&0.108\\\hline\hline
  \end{tabular}
  \end{center}
  \caption{Nose-wise acceptance rates of the proposed BayesEl RJMCMC moves for the gene expression data.}
  \label{tab:accpt}
  \end{table}

In table \ref{tab:accpt} we present the acceptance rates of the RJMCMC moves for the BayesEl regression on various nodes of the gene expression data. Other than DXR with DXPS2 and DXPS3 as parents, all regression has more than $10\%$ acceptance rate. The highest acceptance rate is about $27\%$ achieved for the gene CMK with parents MCT and DXPS3.

\begin{figure}[h]
  \subfigure[Trace-plot of the variance for the regression of PPDS2 \label{fig:sigma}]{
    \resizebox{.45\columnwidth}{!}{\includegraphics{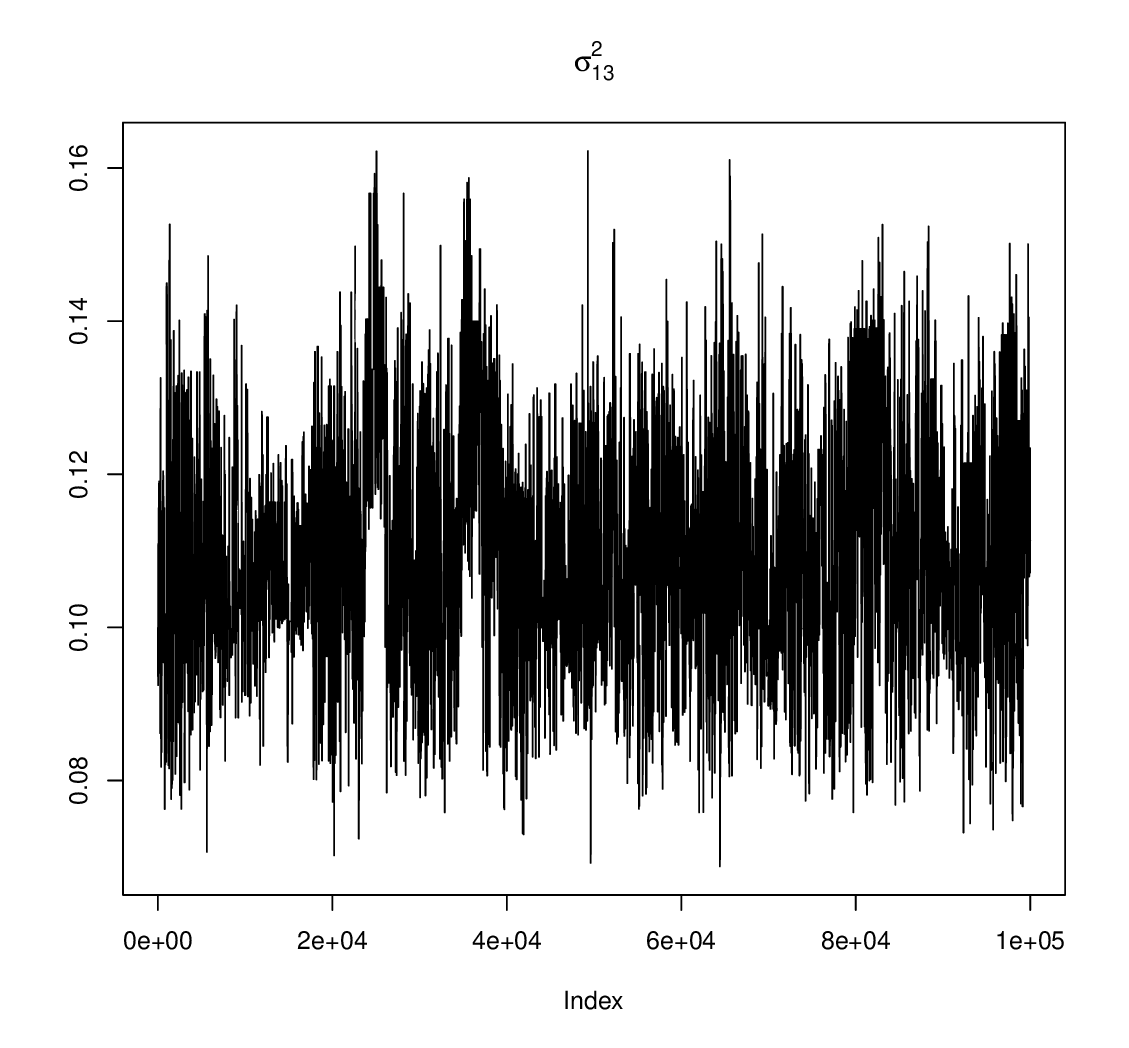}}
  }\quad
   \subfigure[Trace-plot of the regression parameter corresponding to HDS in the regression of PPDS2 \label{fig:hdsPpds2}]{
    \resizebox{.45\columnwidth}{!}{\includegraphics{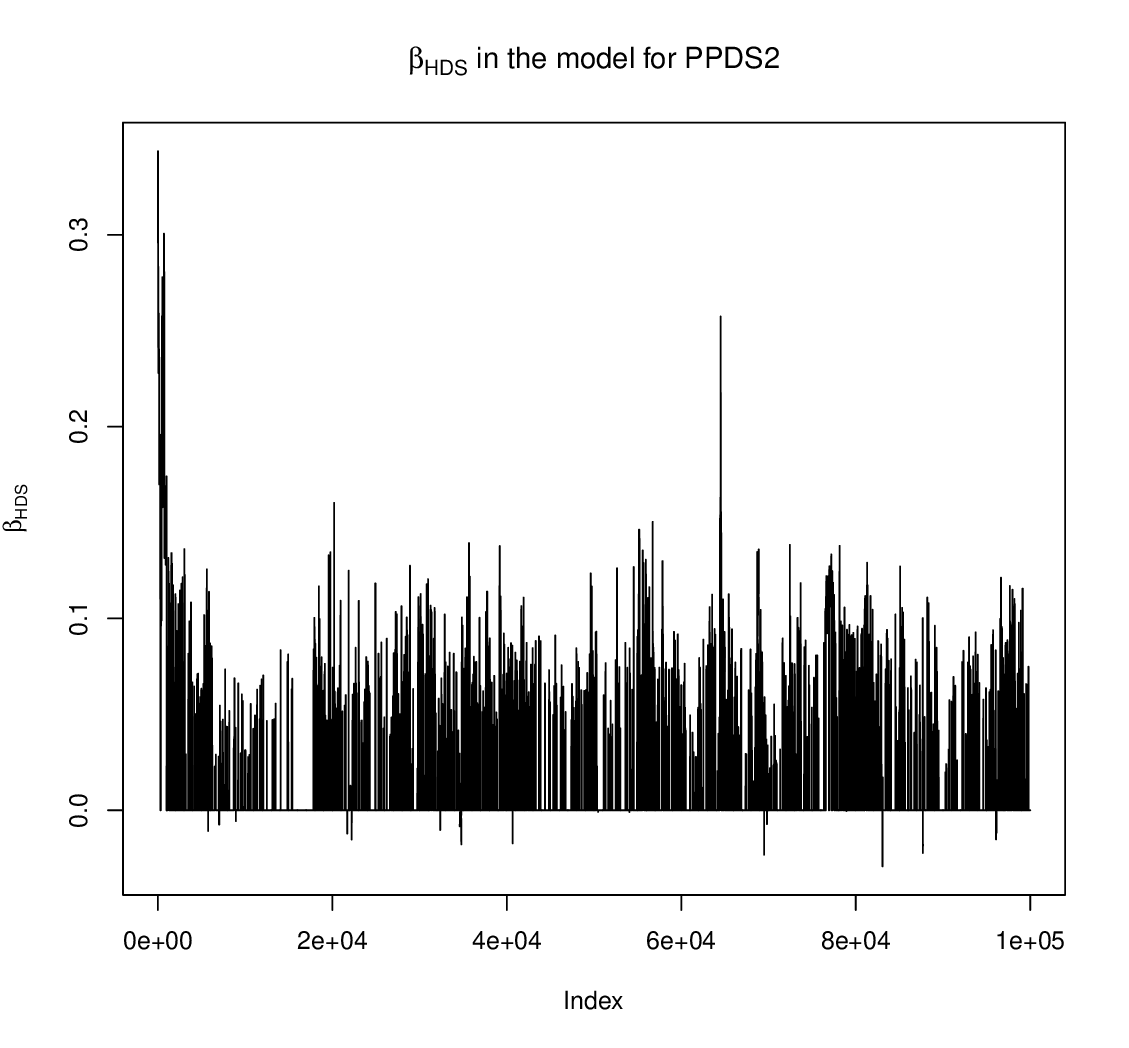}}
   }
    \caption{Some trace-plots of the variables sampled using the proposed BayesEl RJMCMC algorithm}
   \label{fig:traceplots1}
\end{figure}

\begin{figure}[h]
\subfigure[Trace-plot of the regression parameter corresponding to HDS in the regression of GPPS \label{fig:hdsGpps}]{
    \resizebox{.45\columnwidth}{!}{\includegraphics{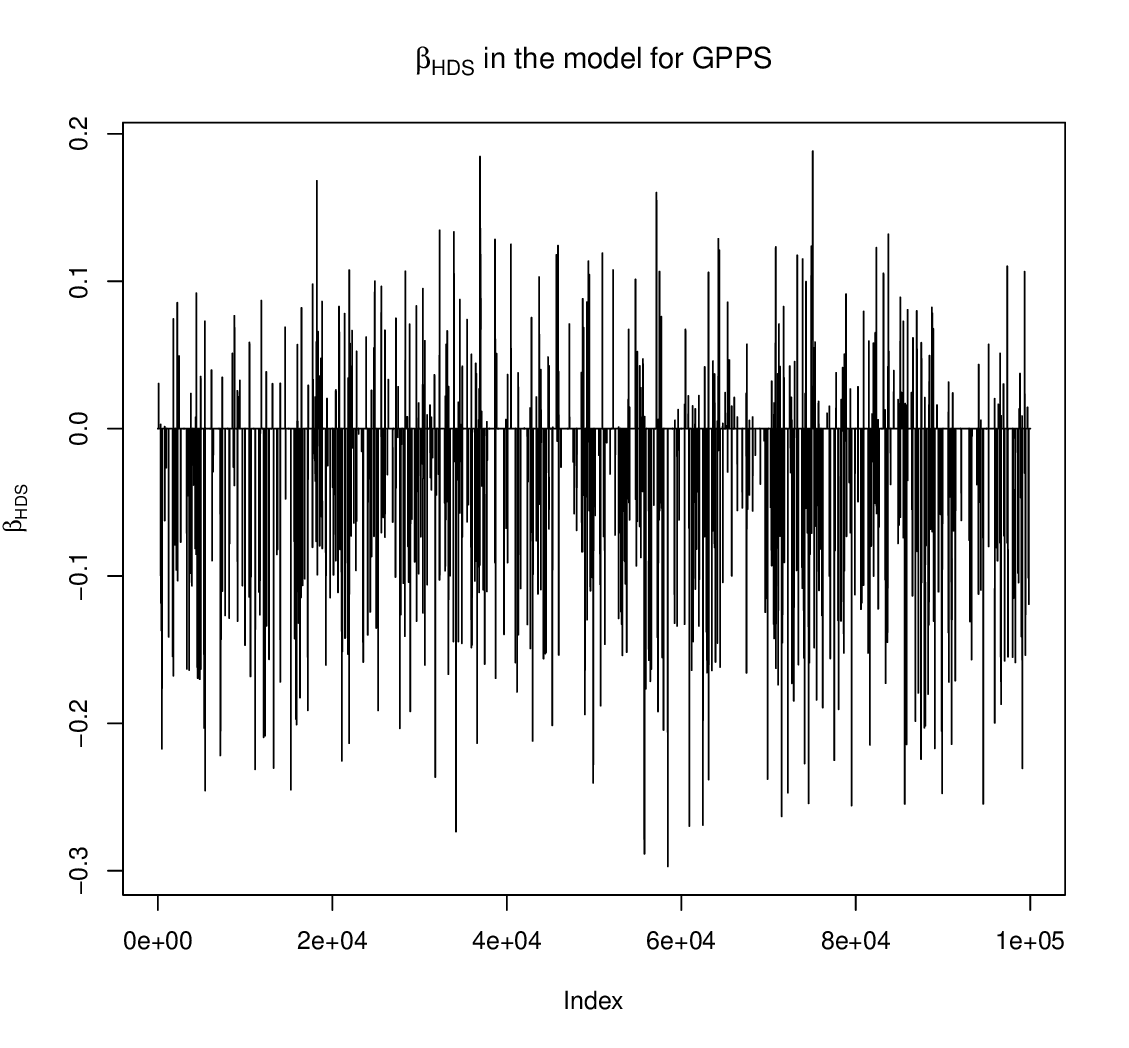}}
  }\quad
   \subfigure[Trace-plot of the regression parameter corresponding to HDR in the regression of GPPS \label{fig:hdrGpps}]{
    \resizebox{.45\columnwidth}{!}{\includegraphics{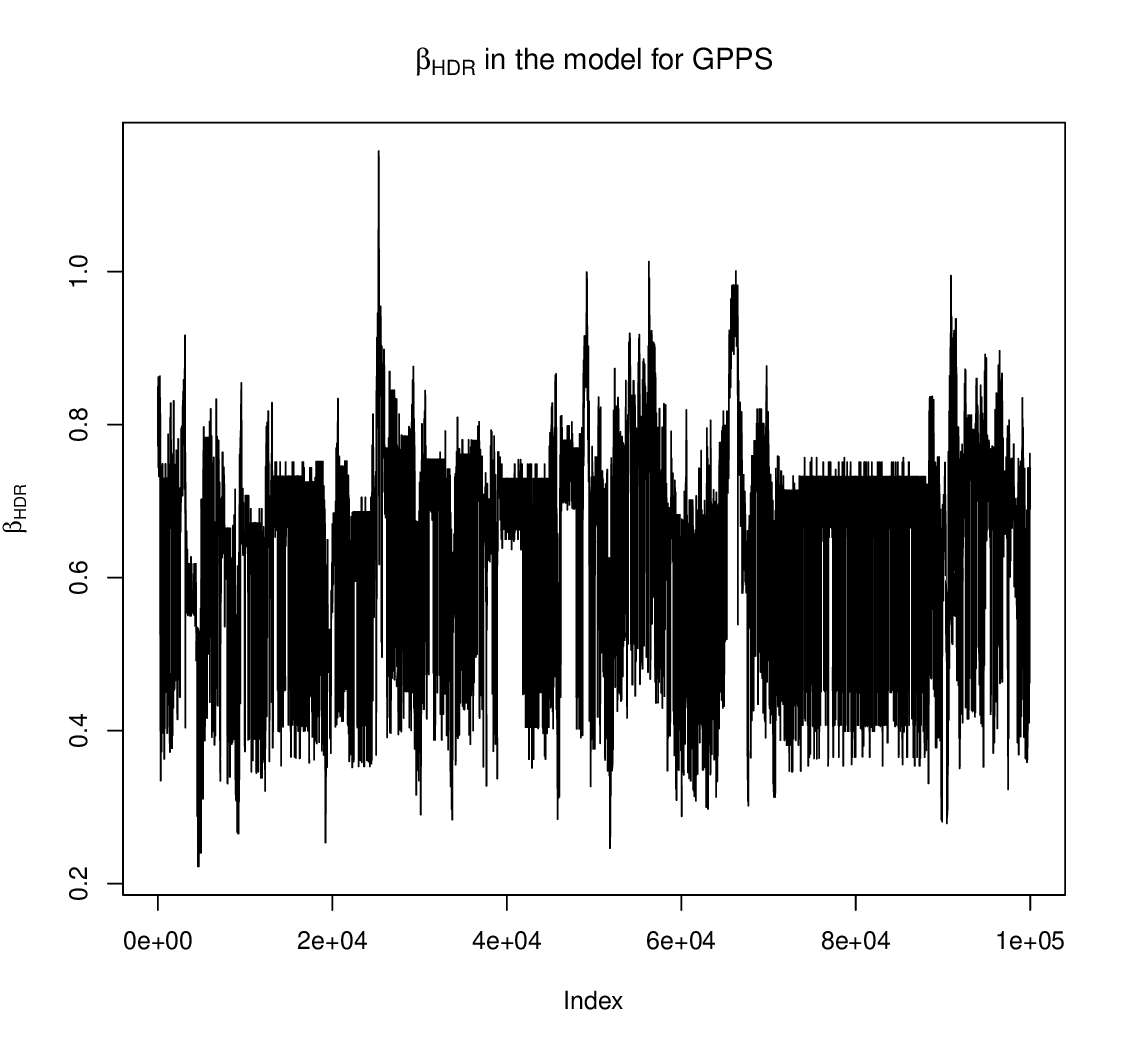}}
   }
   \caption{Some trace-plots of the variables sampled using the proposed BayesEl RJMCMC algorithm}
   \label{fig:traceplots2}
  \end{figure}

In Figures \ref{fig:traceplots1} and \ref{fig:traceplots2} trace-plots of the some regression coefficients and variances are presented.  Figure \ref{fig:sigma} shows the plot of the variance in the regression of gene PPDS2.  The trace-plots of the regression coefficients corresponding to the gene HDS in the set of parents of the genes PPDS2 and GPPS are shown in Figures \ref{fig:hdsPpds2} and \ref{fig:hdsGpps} respectively.
In both cases the genes were not in the parent sets of highest probability.  This is also evident from the figures.  It seems that the chain samples zero very frequently.  The gene HDR is in the set of parents of the gene GPPS.  The trace-plot of the samples drawn can be found in Figure \ref{fig:hdrGpps}.



\section{Data Availability} The datasets used in the article are publicly available.

\bibliographystyle{chicago}
\bibliography{reference}

\end{document}